\documentclass[10pt]{iopart}

\usepackage{tabls}
\usepackage{url}
\usepackage{amsbsy}
\usepackage{amssymb}
\usepackage{graphicx}
\usepackage{color}
\usepackage{hyperref}

\newcommand{\citep}[1]{\cite{#1}}
\newcommand{\citet}[1]{\cite{#1}}
\newcommand{\eqref}[1]{{(\ref{#1})}}
\newcommand{\text}[1]{{#1}}

\begin{document}

\title{Low-frequency gravitational-wave science with eLISA/NGO}

\author{Pau Amaro-Seoane$^{1,13}$,
Sofiane Aoudia$^{1}$,
Stanislav Babak$^{1}$,
Pierre Bin\'etruy$^{2}$,
Emanuele Berti$^{3,4}$,
Alejandro Boh{\'e}$^{5}$,
Chiara Caprini$^{6}$,
Monica Colpi$^{7}$,
Neil J. Cornish$^{8}$,
Karsten Danzmann$^{1}$,
Jean-Fran\c cois Dufaux$^{2}$,
Jonathan Gair$^{9}$,
Oliver Jennrich$^{10}$,
Philippe Jetzer$^{11}$,
Antoine Klein$^{11,8}$,
Ryan N. Lang$^{12}$,
Alberto Lobo$^{13}$,
Tyson Littenberg$^{14,15}$,
Sean T. McWilliams$^{16}$,
Gijs Nelemans$^{17,18,19}$,
Antoine Petiteau$^{2,1}$,
Edward K. Porter$^{2}$,
Bernard F. Schutz$^{1}$,
Alberto Sesana$^{1}$,
Robin Stebbins$^{15}$,
Tim Sumner$^{20}$,
Michele Vallisneri$^{21}$,
Stefano Vitale$^{22}$,
Marta Volonteri$^{23,24}$, and
Henry Ward$^{25}$}

\address{
$^1$Max Planck Inst. f\"ur Gravitationsphysik
(Albert-Einstein-Inst.), Germany\\
$^2$APC, Univ. Paris Diderot, CNRS/IN2P3, CEA/Irfu, Obs. 
    de Paris, Sorbonne Paris Cit{\'e}, France\\
$^{3}$Dept. of Phys. and Astronomy, Univ. of
      Mississippi, University MS 38677, USA\\
$^{4}$Division of Phys., Math., and Astronomy, California 
      Inst. of Tech., Pasadena CA 91125, USA\\
$^5$UPMC-CNRS, UMR7095, Institut d'Astrophysique de Paris, F-75014, 
    Paris, France\\
$^6$Institut de Physique Th{\'e}orique, CEA, IPhT, CNRS, 
    URA 2306, F-91191 Gif/Yvette Cedex, France\\
$^7$Univ. of Milano Bicocca, Milano, I-20100, Italy\\
$^8$Dept. of Phys., Montana State 
   Univ., Bozeman MT 59717, USA\\
$^9$Inst. of Astronomy, Univ. of Cambridge, Madingley Road, 
    Cambridge, UK\\
$^{10}$ESA, Keplerlaan 1, 2200 AG Noordwijk, The Netherlands\\
$^{11}$Inst. of Theoretical Phys. Univ. of Z{\"u}rich,
       8057 Z{\"u}rich Switzerland\\
$^{12}$Washington Univ. St. Louis, St. Louis MO
       63130, USA\\
$^{13}$Institut de Ci{\`e}ncies de l'Espai (CSIC-IEEC), Campus UAB, ES-08193 Bellaterra, Barcelona, Spain\\
$^{14}$Maryland Center for Fundamental Phys., Dept. of Phys., Univ. of Maryland, 
       College Park MD 20742\\       
$^{15}$Gravitational Astrophys. Laboratory, 
       NASA Goddard Spaceflight Center, Greenbelt MD 20771, USA\\
$^{16}$Dept. of Phys., Princeton Univ., Princeton NJ 08544, USA\\
$^{17}$Dept. of Astrophys., Radboud Univ. Nijmegen, The Netherlands\\
$^{18}$Inst. for Astronomy, KU Leuven, 3001 Leuven,
Belgium \\
$^{19}$Nikhef, Science Park 105, 1098 XG Amsterdam, The Netherlands\\
$^{20}$Blackett Lab., Imperial College, London, UK \\
$^{21}$Jet Propulsion Laboratory, California Inst. of Technology, 
       Pasadena CA 91109, USA \\
$^{22}$Univ. of Trento, Dept. of Phys. and INFN, 
       I-38123 Povo, Trento, Italy\\
$^{23}$Institut d'Astrophysique de Paris, 98bis Boulevard Arago, 75014 Paris, France\\
$^{24}$Astronomy Dept., Univ. of Michigan, Ann Arbor MI 48109, USA\\       
$^{25}$Inst. for Gravitational Research, Dept. of Phys. \& Astronomy
       Kelvin Building, Univ. of Glasgow, Glasgow, UK}
\ead{Michele.Vallisneri@jpl.nasa.gov}
1
\newpage
\begin{abstract}
We review the expected science performance of the New Gravitational-Wave Observatory (NGO, a.k.a.\ eLISA), a mission under study by the European Space Agency for launch in the early 2020s. eLISA will survey the low-frequency gravitational-wave sky (from 0.1 mHz to 1 Hz), detecting and characterizing a broad variety of systems and events throughout the Universe, including the coalescences of massive black holes brought together by galaxy mergers; the inspirals of stellar-mass black holes and compact stars into central galactic black holes; several millions of ultracompact binaries, both detached and mass transferring, in the Galaxy; and possibly unforeseen sources such as the relic gravitational-wave radiation from the early Universe.
eLISA's high signal-to-noise measurements will provide new insight into the structure and history of the Universe, and they will test general relativity in its strong-field dynamical regime.
\end{abstract}

\vspace{-16pt}




\section{Introduction}

Over the last two decades, as many as 2,500 articles on space-based gravitational-wave (GW) detection included mentions of LISA (the Laser Interferometer Space Antenna) \cite{lisasciencecase,Jennrich:2009p1398,lisaads}, the space-based GW interferometer planned and developed together by NASA and ESA. This collaboration between the two agencies ended in early 2011 for programmatic and budgetary reasons. In fact, LISA, as brought forth by the entirety of those papers, was more than a space project: it was the concept (and the cherished dream) of a space-based GW observatory that would explore the low-frequency GW sky, in a frequency band ($10^{-4}\mbox{--}1$ Hz) populated by millions of sources in the Galaxy and beyond: compact Galactic binaries; coalescing massive black holes (MBHs) throughout the Universe; the captures of stellar remnants into MBHs; and possibly relic radiation from the early Universe.

All along its evolution, the LISA design remained based on three architectural principles developed and refined since the 1970s: a triangular spacecraft formation with Mkm arms, in Earth-like orbit around the Sun; the continuous monitoring of inter-spacecraft distance oscillations by laser interferometry; drag-free control of the spacecraft around freely falling test masses, the reference endpoints for the distance measurements, achieved using micro-Newton thrusters. The current incarnation of this concept is eLISA (evolved LISA), a mission under consideration by ESA alone (under the official name of NGO, the New Gravitational-wave Observatory) for launch in 2022 within the Cosmic Vision program.

The eLISA design would achieve a great part of the LISA science goals, as presented in \cite{lisasciencecase}, and endorsed by the 2010 U.S.\ astronomy and astrophysics decadal survey \cite{national2010New}. This article reviews eLISA's science performance (sensitivity, event rates, and parameter estimation), as scoped out by these authors in the spring and summer of 2011, and as discussed in full in Ref.\ \cite{2012arXiv1201.3621A}. This article is organized as follows: in Sec.\ \ref{sec:elisa} we provide a very brief overview of eLISA and its GW sensitivity, while later sections are organized by science topics. In Sec.\ \ref{sec:compactbinaries}, we discuss the astrophysics of compact stellar-mass binaries in the Galaxy;
in Sec.\ \ref{sec:massiveblackholes}, the origin and evolution of the massive BHs found at the center of galaxies, as studied through their coalescence GWs; in Sec.\ \ref{sec:emris}, the dynamics and populations of galactic nuclei, as probed through the captures of stellar-mass objects into massive BHs; in Sec.\ \ref{sec:gravity}, the fundamental theory of gravitation, including its behavior in the strong nonlinear regime, its possible deviations from general-relativistic predictions, and the nature of BHs; in Sec.\ \ref{sec:cosmo}, the (potentially new) physics of the early Universe, and the measurement of cosmological parameters with GW events. Last, in Sec.\ \ref{sec:conclusions} we draw our conclusions, and express a wish.

\section{The eLISA mission and sensitivity}
\label{sec:elisa}

We refer the reader to \cite{2012arXiv1201.3621A} for a detailed description of the eLISA architecture. eLISA has a clear LISA heritage, with a few substantial differences. The eLISA arms will be shorter (1 Mkm), simplifying the tracking of distant spacecraft, alleviating requirements on lasers and optics, and reducing the mass of the propellant needed to reach the final spacecraft orbits. The orbits themselves may be slowly drifting away from Earth, again saving propellant, and the nominal mission duration will be two years, extendable to five. As much existing hardware as possible, including the spacecraft bus, will be incorporated from the LISA Pathfinder mission, scheduled for launch by ESA in 2014. The three spacecraft will consist of one ``mother'' and two simpler ``daughters,'' with interferometric measurements along only two arms, for cost and weight savings that make launch possible with smaller rockets than LISA. (Note that LISA was to be built with laser links along the three arms, but it was not a requirement that they would operate throughout the mission.)

\begin{figure}
\flushright
\includegraphics[width=0.7\textwidth]{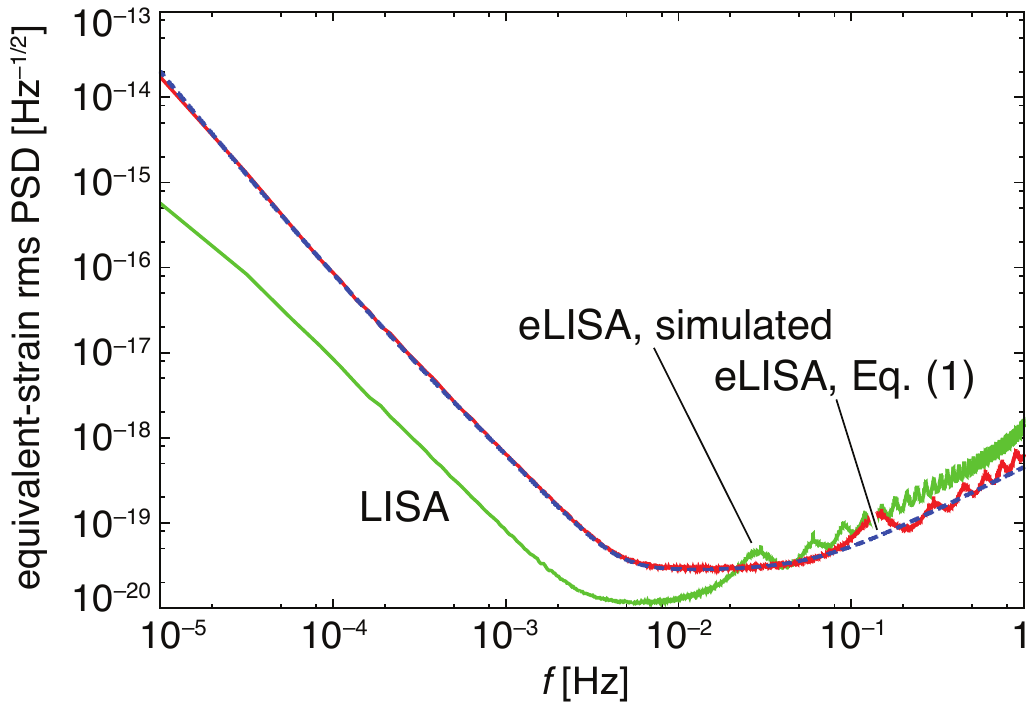}
\caption{eLISA equivalent-strain noise, averaged over source sky location and polarization, as a function of frequency. The solid red curve was obtained with the LISACode 2.0 simulator \citep{petiteau:2008PhRvD..77b3002P}, while the dashed blue curve is plotted from Eq.\ \eqref{eq:sens}. For comparison, the dotted green curve shows the LISA sensitivity.}
\label{fig:sensitivity}
\end{figure}

The eLISA power-spectral-density requirement for the residual test-mass acceleration is  $S_\mathrm{acc}(f) = 2.13 \times 10^{-29} (1 + { 10^{-4}\,{\rm Hz} / f }) \, {\rm m}^2\,{\rm s}^{-4}\,{\rm Hz}^{-1}$, while the position-noise requirement breaks up into 
$S_\mathrm{sn}(f) = 5.25 \times 10^{-23} \; {\rm m}^2\,{\rm Hz}^{-1}$ for shot noise, and 
$S_\mathrm{omn}(f) = 6.28 \times 10^{-23} \; {\rm m}^2\,{\rm Hz}^{-1}$
for all other measurement noises. With these requirements, eLISA achieves the equivalent-strain noise plotted in Fig.\ \ref{fig:sensitivity}, and approximated analytically by
\begin{equation} \fl
\quad \quad \quad
S(f) = \frac{20}{3} \, \frac{4 \, S_{\rm acc}(f) / (2 \pi f)^4 + S_{\rm sn}(f) + S_{\rm omn}(f)}{L^2} \times \biggl(   1 + \Bigl( \frac{f}{0.41 \, c / 2 L} \Bigr) \biggr)^2,
\label{eq:sens}
\end{equation}
where $L = 1$ Mkm, $c$ is the speed of light, and $S(f)$ has already been normalized to account for the sky-averaged eLISA response to GWs.
At the frequency of best sensitivity ($\sim 12$ mHz), the eLISA noise would yield SNR = 1 for a constant-amplitude, monochromatic source of strain $3.6 \times 10^{-24}$ in a two-year measurement.
The requirement on the useful measurement band is $10^{-4}$ Hz to 1 Hz, with a goal of $3 \times 10^{-5}$ Hz to 1 Hz.

\section{Compact binaries in the Galaxy}
\label{sec:compactbinaries}

(See \cite{2009CQGra..26i4030N,marsh:2011CQGra..28i4019M} for deeper reviews.)
The most numerous sources in the low-frequency GW sky observed by eLISA will be short-period binaries of two compact objects such as white dwarfs (WDs) or neutron stars (NSs). These systems have weak GW emission relative to the much heavier massive-BH binaries, but are numerous in the Galaxy and even in the Solar neighborhood. To date, astronomers have observed about 50 ultra-compact binaries with periods shorter than one hour, comprising both detached systems and interacting binaries where mass is being transferred from one star to the other. Wide-field and synoptic surveys such as SDSS and PTF (and in the future, PanSTARRS, EGAPS, and LSST) will continue to enlarge this sample \citep{rau:2010ApJ...708..456R,levitan:2011ApJ...739...68L}. Interacting ultra-compact binaries with NS accretors are found by all-sky X-ray monitors and in dedicated surveys \citep{jonker:2011ApJS..194...18J}.

A large subset of known systems will be guaranteed \textbf{verification sources} for eLISA \citep{stroeer:2006:lvb}; their well-modeled GW signals will be detected within the first few weeks to months of operation, verifying instrument performance. The most promising verification binaries are the shortest-known-period interacting systems HM Cnc (with a period of 5.4 min \citep{roelofs:2010:se}), V407 Vul ($P = 9.5$ min \citep{2006ApJ...649..382S}), and ES Cet \citep{2011MNRAS.413.3068C} and the recently discovered detached system SDSS J0651+28 ($P = 12$ min \citep{brown:2011ApJ...737L..23B}).
\begin{figure}
  \flushright
  \includegraphics[width=0.8\textwidth]{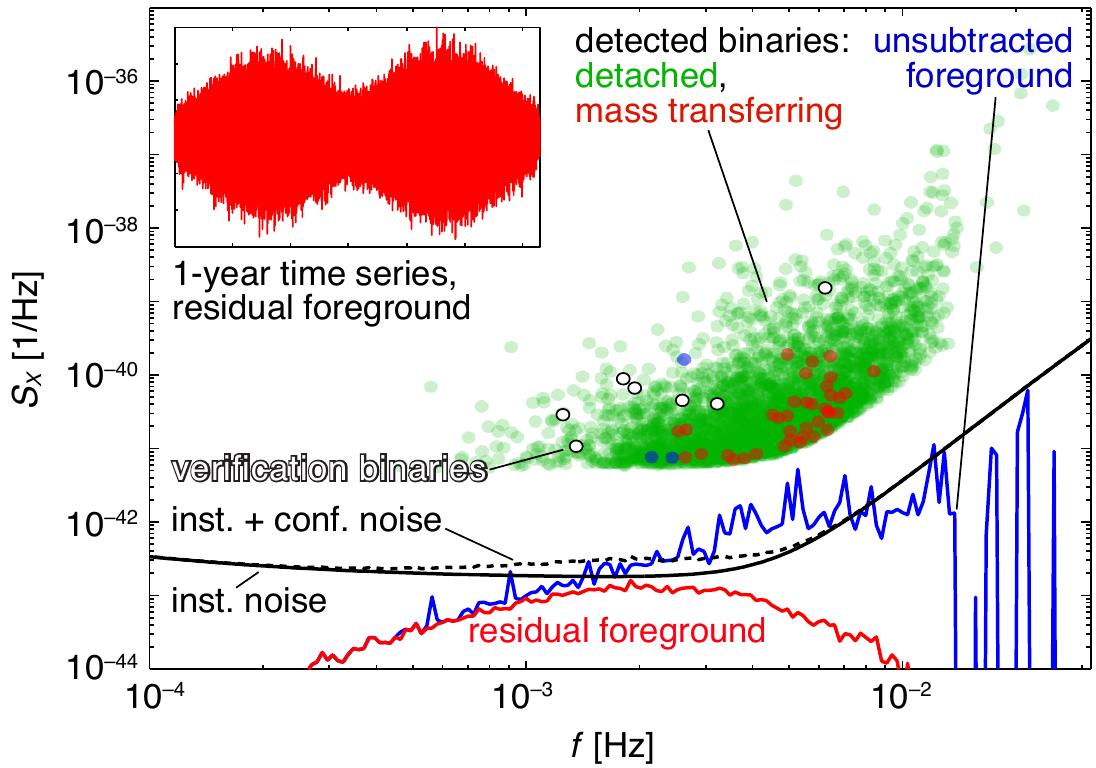}
  \caption{\textbf{Main figure}: power spectral density of the stochastic GW foreground from Galactic binaries, \emph{before} (blue) and \emph{after} (red) the subtraction of individually resolvable systems, which are plotted as green and red/blue dots (for detached and mass-transferring systems). A few known verification binaries are shown as white dots. The solid/dashed black curves trace instrument noise alone/with confusion noise. Spectra are shown for the observable ``$X$'' of Time Delay Interferometry (see, e.g., \cite{PhysRevD.72.042003}); subtraction is simulated for a two-year observation and threshold $\mathrm{SNR} = 7$; resolvable systems are placed a factor $\mathrm{SNR}^2$ above the combined instrument and confusion noise. \textbf{Inset}: time series of the residual foreground, which carries information about the number and distribution of binaries in the Galaxy.\label{fig:binaries}}
  \vspace{-8pt}
\end{figure}

eLISA will individually detect and determine the periods of \textbf{several thousand currently unknown compact binaries} (in our estimate, 3,500--4,100 systems for a two-year observation; \cite{2012arXiv1201.3621A,2012arXiv1201.4613N}), while the combined signals of tens of millions unresolvable systems will form a \textbf{stochastic GW foreground} at frequencies below a few mHz (\cite{yu:2010:gw,ruiter:2010ApJ...717.1006R}; see Fig.\ \ref{fig:binaries}.) About $\sim 500$ close or high-frequency ($> 10$ mHz) sources will be seen with large SNRs, allowing the determination of sky position to better than 10 $\mathrm{deg}^2$, of frequency derivative to 10\%, of inclination to 10 deg, and of distance to 10\%. This large sample will allow a detailed study of the Galactic population, which is poorly constrained by EM observations and theoretical predictions \citep{roelofs:2007:sdss}.

Detections will be dominated by \textbf{double WD binaries} with the shortest periods (5--10 minutes). Their mergers are candidate progenitors for many interesting systems: type Ia \citep{pakmor:2010:sltIa} and peculiar supernovae \citep{perets:2010Natur.465..322P,waldman:2011ApJ...738...21W}; single subdwarf O and B stars, R Corona Borealis stars and maybe all massive WDs \citep{webbink:1984:dwd}; and possibly the rapidly spinning NSs observed as ms radio pulsars and magnetars \cite{levan:2006:grb}. These binaries are short lived, very faint for telescopes, and scarce (few thousand in the whole Galaxy), so GWs will provide a unique window on their physics. eLISA will determine their merger rate, constrain their formation, and illuminate the preceding phases of binary evolution, most notably the common-envelope phase.

\textbf{Common-envelope evolution} is crucial to most binary systems that produce high-energy phenomena such as $\gamma$-ray bursts and X-ray emission, but our understanding of its physics and outcome is limited \citep{{taam:2000:cee,taam:2010NewAR..54...65T}} and challenged by observations \citep{nelemans:2005MNRAS.356..753N,demarco:2011MNRAS.411.2277D}. The standard scenario is as follows. Most stars in the Universe are in binaries, and roughly half of binaries are formed at close enough separations that the stars will interact as they evolve into giants or supergiants.  Following runaway mass transfer, the companion of the giant can end up inside the outer layers (the envelope) of the giant; dynamical friction reduces the velocity of the companion, shrinking the orbit and transferring angular momentum and energy into the envelope; the envelope eventually becomes unbound, leading to a very compact binary consisting of the core of the giant and the original companion \citep{paczynski:1976:secbs}. 

eLISA will also test dynamical interactions in \textbf{globular clusters}, which produce an overabundance of ultra-compact X-ray binaries consisting of a NS accreting material from a WD companion. The eLISA angular resolution will be sufficient to distinguish WD binaries in clusters, verifying whether they are also plentiful.

The eLISA measurements of individual short-period binaries will provide a wealth of information on the physics of tidal interactions and the stability of mass transfer. For detached systems with little or no interaction, the evolution of the GW signal is dominated by gravitational radiation:
\begin{equation}
  \label{eq:fders}
  h \propto {\cal M}^{5/3} f^{2/3} D^{-1}, \quad 
  \dot{f} \propto {\cal M}^{5/3} f^{11/3}, \quad
  \ddot{f} = \frac{11}{3} \frac{\dot f}{f},
\end{equation}
where $h$ is the GW strain, $f$ the GW frequency, $\mathcal{M} = (m_1 m_2)^{3/5}/(m_1+m_2)^{1/5}$ is the chirp mass with $m_1$, $m_2$ the individual masses, and $D$ is the distance. Thus, measuring $h$, $f$, and $\dot{f}$ (which will be possible in 25\% of systems) provides $\mathcal{M}$ and $D$; measuring also $\ddot{f}$ (which may be possible for a few high-SNR systems) tests secular effects from tidal and mass-transfer interactions. Short-term variations are not likely to prevent detection \citep{stroeer:2009:stv}, and the precision of $\dot{f}$ and $\ddot{f}$ determination increases with the duration of the mission.

\textbf{Tidal interactions} are possible when at least one binary component does not corotate with the orbital motion, or when the orbit is eccentric. Their strength is unknown \citep{marsh:2004:mtwd}, and has important consequences on the tidal heating (and possibly optical observability) of WD binaries, as well as the stability of \textbf{mass transfer}. This process begins after gravitational radiation shrinks detached binaries to sufficiently close orbits (with $P \sim$ a few minutes) that one of the stars fills its Roche lobe and its material can leak to the companion. Mass transfer can be self-limiting, stable, or unstable, depending on the resulting evolution of the orbit and of the donor radius. Unstable transfer leads to mergers; stable systems (the interacting WD binaries known as AM CVn systems, as well as ultra-compact X-ray binaries) will be observed -- and counted -- by eLISA in the early stages of mass transfer \citep{marsh:2011CQGra..28i4019M}. Efficient tidal coupling can return angular momentum from the accreted material to the orbit \citep{marsh:2004:mtwd,dsouza:2006:dmt,racine:2007:ndts}, slowing the inspiral and increasing the fraction of WD binaries that survive the onset of mass transfer from 0.2\% to 20\% \citep{nelemans:2001:ps2}.

The \textbf{unresolved foreground} from Galactic binaries will provide an additional noise component for the detection of loud broadband signals (see the dashed line in Fig.\ \ref{fig:binaries}), but it also contains precious astrophysical information. Its overall level measures the total number of binaries (mostly double WDs); its spectral shape characterizes their history and evolution; and its yearly modulation \cite{edlund:2005:wdw}, together with the distance determinations from many individual systems, constrains the distribution of sources in the different Galactic components. Thus eLISA will probe dynamical effects in the Galactic center, which may increase the number of tight binaries \citep{alexander:2005:spmbh}; it will measure the poorly known scale height of the disk; and it will sample the population of the halo \citep{ruiter:2009:hwdb,yu:2010:gw}, which hosts two anomalous AM CVn systems and which may have a rather different compact-binary population than the rest of the Galaxy.
Furthermore, the eLISA measurements of orbital inclinations for individual binaries, compared with the overall angular momentum of the Galaxy, will provide hints on the formation of binaries from interstellar clouds.

eLISA will also constrain the formation rate and numbers of \textbf{NS binaries} and \textbf{ultra-compact stellar-mass BH binaries}, throughout the Galaxy and without EM selection effects. These numbers are highly uncertain, but as many as several tens of systems may be detectable by eLISA \citep{nelemans:2001:ps2,belczynski:2010:dco}, complementing the ground-based GW observations of these same systems in other galaxies (and at much shorter periods). 

More generally, the astrophysical populations and parameters probed by eLISA will be different from, and complementary to, what can be deduced from EM observations. For instance, eLISA will be sensitive to binaries at the Galactic center and throughout the Galaxy, while Gaia \cite{2001A&A...369..339P} will be limited to the Solar neighborhood; GWs encode distances and orbital inclinations, while EM emission is sensitive to surface processes. Dedicated observing programs and public data releases will allow simultaneous and follow-up EM observations of binaries identified by eLISA.

\section{Massive black-hole binaries}
\label{sec:massiveblackholes}

(See \cite{2012arXiv1201.3621A} for a much deeper review.)
According to the accretion paradigm \citep{Salpeter:1964,zeldovich:1964:rgw,krolik:1999:agn}, supermassive BHs of $10^6\mbox{--}10^9 \, M_\odot$ power quasars---active galactic nuclei so luminous that they often outshine their galaxy host, which are detected over the entire cosmic time accessible to our telescopes. \emph{Quiet} supermassive BHs are ubiquitous in our low-redshift Universe, where they are observed to have masses closely correlated with key properties of their galactic host (see \citep{gultekin:2009:ms}, and refs.\ therein)
leading to the notion that galaxies and their nuclear MBHs form and evolve in symbiosis
(see, e.g., \cite{dimatteo:2005:eiq,hopkins:2006:umm,croton:2006MNRAS.365...11C}).

In the currently favored cosmological paradigm, regions of higher-density cold dark matter in the early Universe form self-gravitating halos, which grow through mergers with other halos and accretion of surrounding matter; baryons and MBHs are thought to follow a similar bottom-up \emph{hierarchical clustering} process \citep{white:1978MNRAS.183..341W,haiman:1998ApJ...503..505H,haehnelt:1998MNRAS.300..817H,wyithe:2002ApJ...581..886W,volonteri:2003:amh}. 
MBHs may be born as \emph{small seeds} ($10^2 \mbox{--} 10^3 \, M_\odot$) from the core collapse of the first generation of ``Pop III'' stars formed from gas clouds in light halos at $z \sim 15\mbox{--}20$ \citep{madau:2001:mbhIII,volonteri:2003:amh}; or as \emph{large seeds} ($10^3 \mbox{--} 10^5 \, M_\odot$) from the collapse of very massive quasi-stars formed in much heavier halos at $z \sim 10\mbox{--}15$ \citep{haehnelt:1993MNRAS.263..168H,loeb:1994:cbg}; or by runaway collisions in star clusters \citep{devecchi:2009ApJ...694..302D}; or again by direct gas collapse in mergers \cite{mayer:2010Natur.466.1082M} (See \citep{volonteri:2010A&ARv..18..279V,2012AdAst2012E..12S} and refs.\ therein).
The seeds then evolve over cosmic time through intermittent, copious accretion and through mergers with other MBHs after the merger of their galaxies.

The cosmic X-ray background from active MBHs at $z < 3$ suggests that radiatively efficient accretion played a large part in building up MBH mass \citep{marconi:2004:lsmbh,yu:2002:oc,soltan:1982:moq}, so information about the initial mass distribution is not readily accessible in the local Universe. By contrast, eLISA will measure the masses of the original seeds from their merger events. Furthermore, it is unknown \cite{volonteri:2007:bhs} whether accretion proceeds \emph{coherently} from a geometrically thin, corotating disk \citep{shakura:1973A&A....24..337S} (which can spin MBHs up to the $J/M^2 = 0.93\mbox{--}0.99$ limit imposed by basic physics \cite{thorne:1974:dabh,gammie:2004:bhse}) or \emph{chaotically} from randomly oriented episodes \citep{king:2006MNRAS.373L..90K} (which typically result in smaller spins). eLISA's accurate measurements of MBH spins will provide evidence for either mechanism \cite{berti:2008:cbhse}.

After a galactic merger, the central MBHs spiral inward, together with their bulge or disc, under the action of dynamical friction, and \emph{pair} as a pc-scale Keplerian binary \citep{begelman:1980Natur.287..307B,chandrasekhar:1943:dfI,ostriker:1999ApJ...513..252O,colpi:1999ApJ...525..720C,mayer:2007Sci...316.1874M}; MBH binaries are then thought to \emph{harden} into gravitational-radiation--dominated systems by ejecting nearby stars (assuming a sufficient supply) 
\citep{quinlan:1996:dembhb,khan:2011ApJ...732...89K,preto:2011ApJ...732L..26P}
or by gas torques and flows in gas-rich environments \citep{escala:2004:rgm,dotti:2007:smbh,cuadra:2009MNRAS.393.1423C}; the final binary \emph{coalescence} is the most luminous event in the Universe (albeit in GWs). BH mergers have been explored only recently by numerical relativity \citep{2011arXiv1107.2819S}, showing how the mass and spin of the final BH remnant arise from those of the binary components, and predicting remarkable physical phenomena such as large remnant recoils for peculiar spin configurations \citep{2011PhRvL.107w1102L}. The predicted coalescence rate in the eLISA frequency band ranges from a handful up to few hundred events per year, depending on theoretical assumptions (\citep{haehnelt:1994:lfg,wyithe:2003:lfg,sesana:2004:lfg,enoki:2004:gws,sesana:2005:gws,rhook:2005:rer,koushiappas:2006:tms,sesana:2007:imprint}).

eLISA will be sensitive to GW signals from all three phases of MBH coalescence (inspiral, merger, and ring-down \citep{flanagan:1998:mgwa}). To assess the eLISA science performance in this area, after experimenting with different waveform families, we modeled these signals with the  ``PhenomC'' phenomenological waveforms \citet{santamaria:2010PhRvD..82f4016S}, which stitch together post-Newtonian (PN) inspiral waves \citep{blanchet:2006:grp} with frequency-domain fits to numerically modeled late-inspiral and ringdown waves.%
\begin{figure}
  \includegraphics[width=\textwidth]{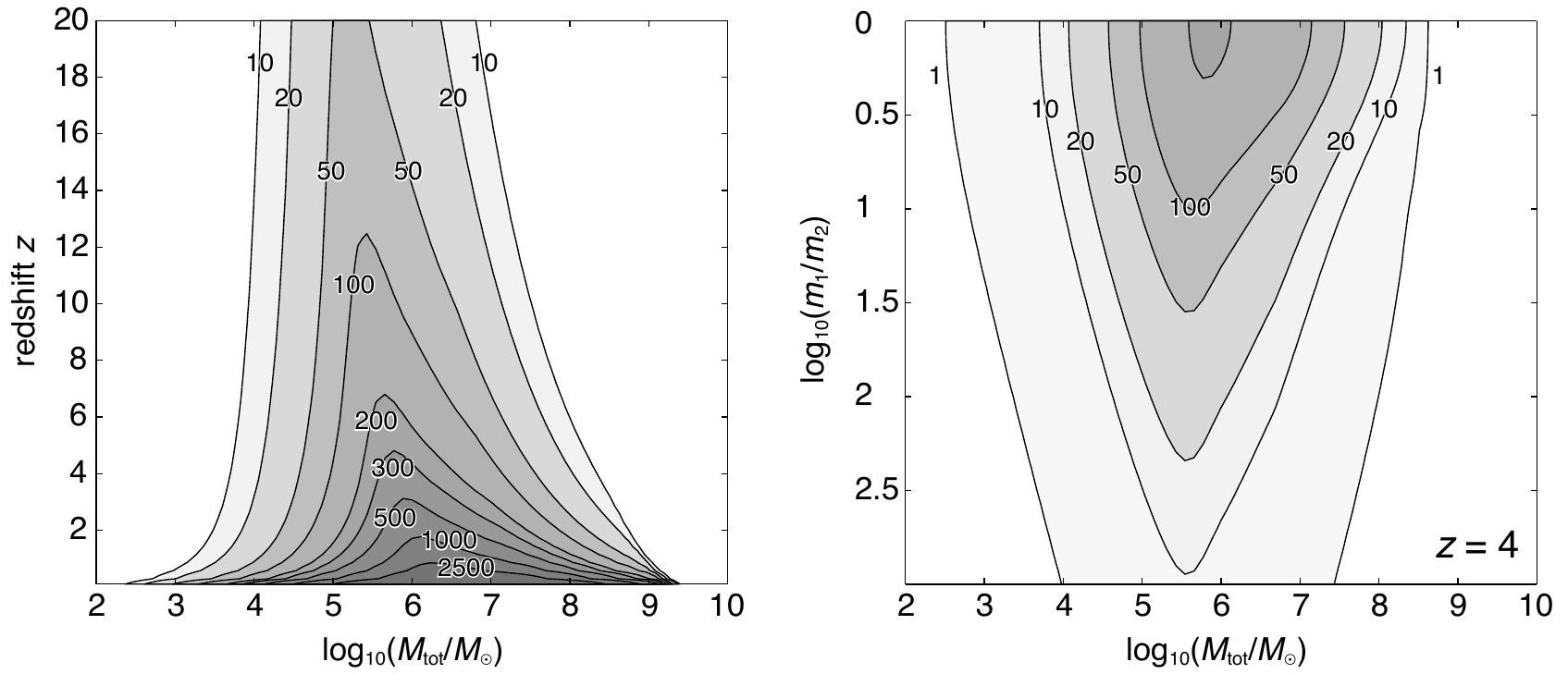}
  \caption{\textbf{Left}: constant-level contours of sky- and polarization-averaged SNR for equal-mass non-spinning binaries as a function of total rest mass $M_\mathrm{tot}$ and cosmological redshift $z$. The SNR includes inspiral, merger and ringdown. \textbf{Right}: SNR contours as a function of $M_\mathrm{tot}$ and mass ratio $q = m_1/m_2$.\label{fig:mbhSNR}}
  \vspace{-8pt}
\end{figure}

The first metric of performance is the \textbf{detection SNR}, angle-averaged over sky position and source orientation, which is plotted in Fig.\ \ref{fig:mbhSNR} as a function of total rest mass and cosmological redshift (left panel) and as a function of total rest mass and mass ratio for binaries at $z = 4$ (right panel). eLISA covers almost all the mass--redshift parameter space of MBH astrophysics: any equal-mass binary with $M_\mathrm{tot} = 10^4\mbox{--}10^7 \, M_\odot$ (the crucial ``middleweight'' range inaccessible to EM observations beyond the local Universe) can be detected (with $\mathrm{SNR} > 10$) out to the highest redshifts, while equal-mass binaries with $M_\mathrm{tot} > 10^5 \, M_\odot$ are seen in detail as strong signals ($\mathrm{SNR} > 100$) out to $z = 5$. Binaries with $M_\mathrm{tot} > 10^5 \, M_\odot$ and mass ratios $\lesssim 10$ are seen with $\mathrm{SNR} > 20$ out to $z = 4$.

\begin{figure}
  \flushright
  \includegraphics[width=\textwidth]{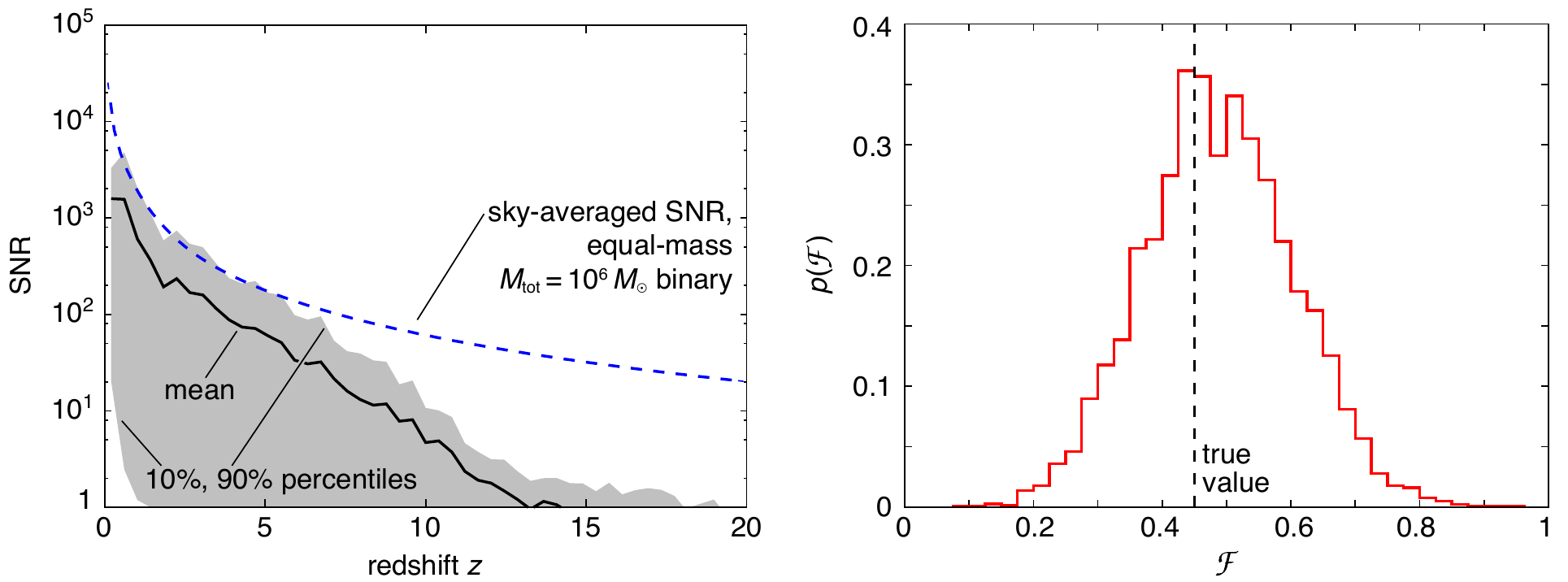}
  \caption{\textbf{Left}: distribution of expected SNR for MBH mergers as a function of $z$, computed from the \textbf{SE}/\textbf{LE}/\textbf{SC}/\textbf{LC} metacatalog (see main text). \textbf{Right}: likelihood for the mixing fraction $\mathcal{F}$, for an individual realization of mixed model $\mathcal{F}\, \mathbf{SE}+(1-\mathcal{F})\mathbf{LE}$ with $\mathcal{F}=0.45$ (see main text).\label{fig:mbhSNRz}}
\end{figure}

To evaluate expected SNRs in the context of \textbf{realistic MBH populations}, we consider four fiducial scenarios (\textbf{SE}, \textbf{LE}, \textbf{SC}, \textbf{LC}) where MBHs originally form from \textbf{S}mall ($\sim 100 \, M_\odot$) or \textbf{L}arge seeds ($\sim 10^5 \, M_\odot$), and where they subsequently grow by \textbf{E}xtended or \textbf{C}haotic accretion. (See \citep{arun:2009:petf} for details; here we enhance that analysis by including random spin--orbit misalignments up to 20 deg in \textbf{E} models \citep{dotti:2010MNRAS.402..682D}). For each scenario we generate multiple catalogs of merger events, and join them in equal proportions into a single metacatalog. Figure \ref{fig:mbhSNRz} shows the resulting distribution of SNR with $z$: eLISA will
detect sources with $\mathrm{SNR} \gtrsim 10$ out to $z \lesssim 10$, a limit imposed by masses of the expected binary population as a function of $z$.
\begin{figure}
  \flushright
  \includegraphics[width=0.8\textwidth]{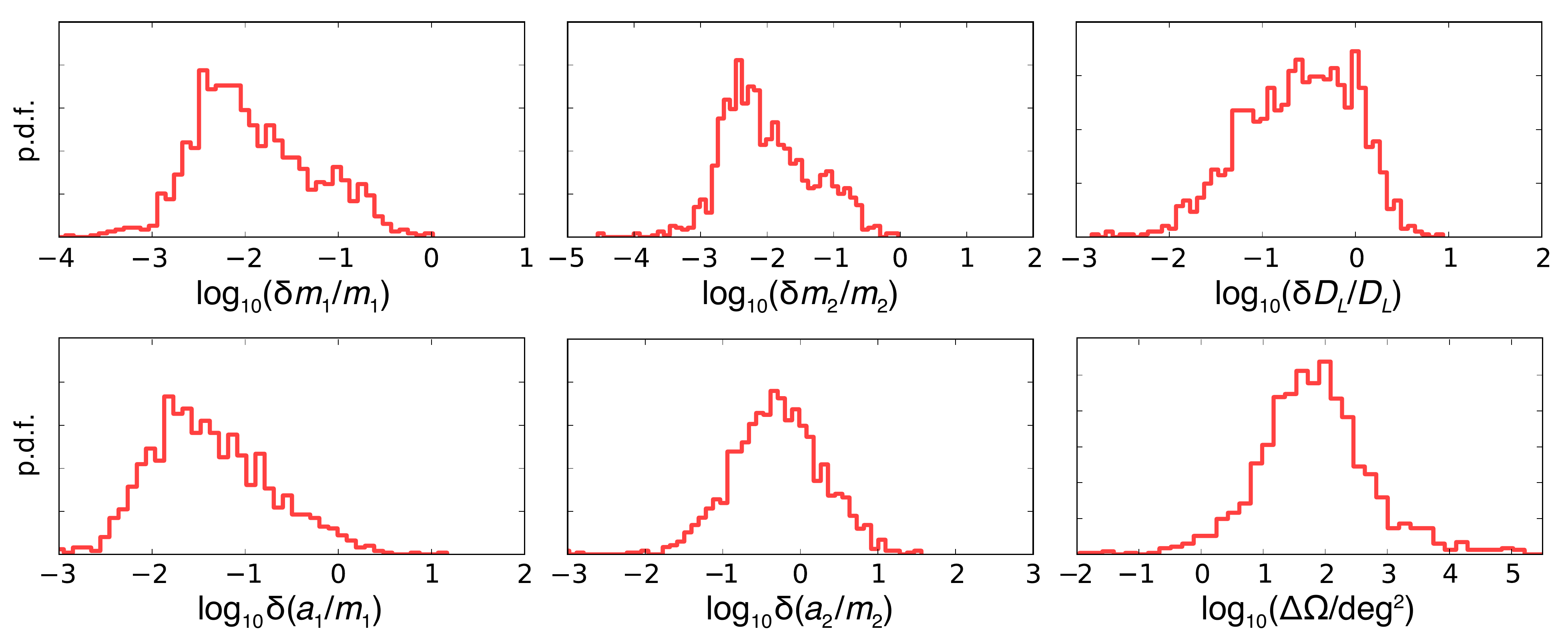}
  \caption{Parameter-estimation accuracy (relative frequency of fractional or ab\-so\-lute errors over \textbf{SE}/\textbf{LE}/\textbf{SC}/\textbf{LC} metacatalog) for primary and secondary \emph{redshifted} MBH masses and dimensionless spins ($m_1$ and $m_2$, $a_1/m_1$ and $a_2/m_2$, respectively), luminosity distance $D_L$ and sky position $\Delta\Omega$.\label{fig:mbhparest}}
\vspace{-8pt}
\end{figure}

For the same metacatalog, Fig.\ \ref{fig:mbhparest} shows the \textbf{expected accuracy of parameter determination}, estimated using a Fisher-matrix approach based on PN inspiral waveforms with spin-induced precession, augmented with PhenomC merger--ringdown waveforms to account for the final ``hang up'' behavior driven by the spin components aligned with the orbital angular momentum.
eLISA can determine the \emph{redshifted} component masses ($m_\mathrm{redshift} = (1+z) \, m_\mathrm{rest}$) to $0.1\mbox{--}1$ \%; the primary-MBH spins to 0.01--0.1; and the secondary-MBH spins to 0.1 in a fraction of systems. (Compare with EM MBH-mass uncertainties $\sim 15\mbox{--}200$\%, except for the Milky Way MBH, and with very large MBH-spin uncertainties from K$\alpha$ iron line fits \cite{McClintock:2011zq}.)
The errors in $D_L$ have a wider spread, from a few percent to virtual non-determination, while sky position $\Omega$ is typically determined to 10--1000 $\mathrm{deg}^2$. Compared to previous published estimates for LISA, the accuracy in determining both $D_L$ and $\Omega$ is reduced for eLISA by having interferometric measurements only along two arms (although three arms were always a goal, not a requirement, for LISA).

The next order of analysis is to combine multiple MBH-coalescence observations, resulting in a catalog of binary/remnant parameters, into a single \textbf{inference about the mechanisms of MBH formation and evolution} throughout cosmic history. This problem was analyzed extensively by Sesana and colleagues \citep{sesana:2011:rmbh} in the context of LISA. We repeated their analysis for eLISA, by generating 1,000 catalogs of detected mergers (over two years) for each of the four \textbf{SE}/\textbf{LE}/\textbf{SC}/\textbf{LC} scenarios, and comparing the relative likelihood $p(A\,\mathrm{vs.}\,B) = p(A|C) / [p(A|C) + p(B|C)]$ for each pair of scenarios $(A,B)$, for $C = A$ or $B$. We considered only detections with $\mathrm{SNR} > 8$, and used spinless, restricted PN waveforms. Table \ref{tab:odds} shows our results for a relative likelihood threshold 0.95: for instance, the first row on the left shows that if \textbf{SE} is true, it \emph{could be discriminated} from \textbf{LE} and \textbf{LC} in 99\% of realizations, but from \textbf{SC} only in 48\% of realizations; the last row on the left shows that \textbf{LC} \emph{could not be ruled out} in 2\% of realizations when \textbf{SE} or \textbf{SC} are true, but in 22\% of realizations when \textbf{LE} is true. This degeneracy between accretion mechanisms is an artifact of the spin-less assumption; including information about the spin of the final merged MBH, which can be measured in 30\% of detections by way of quasinormal-mode ``spectroscopy'' \cite{berti:2006:gws}, provides essentially perfect discrimination.
\begin{table}
\caption{Model discrimination with eLISA MBH-binary observations. The upper-right half of each table shows the fraction of realizations in which the \emph{row} model would be chosen over the \emph{column} model with a likelihood threshold $> 0.95$, when the \emph{row} model is true. The lower-left half of each table shows the fraction of realizations in which the \emph{row} model cannot be ruled out against the \emph{column} model when the \emph{column} model is true. In the left table we consider only the measured masses and redshift for observed events; in the right table we include also the observed distribution of remnant spins.\label{tab:odds}}
\small \flushright
\begin{tabular}{lccccclcccc}
\hline \hline
&\multicolumn{4}{c}{without spins}&&
&\multicolumn{4}{c}{with spins}\\
& \textbf{SE} & \textbf{SC} & \textbf{LE} & \textbf{LC}  & & & \textbf{SE} & \textbf{SC} & \textbf{LE} & \textbf{LC} \\
\hline
\textbf{SE} & $\times$ & 0.48     & 0.99     & 0.99     && \textbf{SE} & $\times$ & 0.96     & 0.99     & 0.99     \\
\textbf{SC} & 0.53     & $\times$ & 1.00     & 1.00     && \textbf{SC} & 0.13     & $\times$ & 1.00     & 1.00     \\
\textbf{LE} & 0.01     & 0.01     & $\times$ & 0.79     && \textbf{LE} & 0.01     & 0.01     & $\times$ & 0.97     \\
\textbf{LC} & 0.02     & 0.02     & 0.22     & $\times$ && \textbf{LC} & 0.02     & 0.02     & 0.06     & $\times$ \\
\hline \hline
\end{tabular}
\vspace{-8pt}
\end{table}

Last, because no theoretical model will exactly capture the ``true'' formation and evolution history of MBHs, we investigated eLISA's ability of measuring the \emph{mixing fraction} $0 < \mathcal{F} < 1$ in a \textbf{mixture model} $\mathcal{F} A + (1-\mathcal{F}) B$ that produces coalescence events with probability $\mathcal{F}$ from scenario $A$, and $1 - \mathcal{F}$ from $B$. For instance, for the case $\mathcal{F} \, \mathbf{SE} + (1-\mathcal{F}) \mathbf{LE}$ with $\mathcal{F} = 0.45$, $\mathcal{F}$ can be measured with an uncertainty of 0.1 (see right panel of Fig.\ \ref{fig:mbhSNRz}). Although highly idealized, this example shows the potential of eLISA's observations to constrain MBH astrophysics along their entire cosmic history, in mass and redshift ranges inaccessible to EM astronomy.

In closing this section, we note that eLISA may also detect coalescences of BHs with masses of $10^2\mbox{--}10^4 \, M_\odot$ (intermediate-mass BHs, or IMBHs). These events do not result from hierarchical galaxy mergers, but they occur locally under the extreme conditions of star clusters. IMBHs may form in young clusters by way of mass segregation followed by runaway mergers \citep{portegieszwart:2000ApJ...528L..17P,guerkan:2004ApJ...604..632G,portegieszwart:2004Natur.428..724P,freitag:2006JPhCS..54..252F,freitag:2006MNRAS.368..121F}; IMBH binaries may form \emph{in situ} \citep{guerkan:2006ApJ...640L..39G}, or after the collision of two clusters \citep{amaro-seoane:2006ApJ...653L..53A,amaro-seoane:2010:mbhb}. Although the evidence for IMBHs is tentative \cite{miller:2004:imbh,miller:2009CQGra..26i4031M}, eLISA may observe as many as a few coalescences per year \citep{amaro-seoane:2006ApJ...653L..53A} out to a few Gpc \citep{santamaria:2010PhRvD..82f4016S}; it may also detect stellar-mass BHs plunging into IMBHs in the local Universe 
\citep{konstatantinidis:2011arXiv1108.5175K}.

\section{Extreme-mass-ratio inspirals and the astrophysics of dense stellar systems}
\label{sec:emris}

There is of course one galactic nucleus, our own, that can be studied and imaged in great detail \citep{schoedel:2003:sdca,ghez:2003:fmsl,eisenhauer:2005:sinfoni,ghez:2005:sogc,ghez:2008:mdp,gillessen:2009:mso}. The central few parsecs of the Milky Way host a dense, luminous star cluster centered around the extremely compact radio source SgrA$^*$. The increase in stellar velocities toward SgrA$^*$ indicates the presence of a $(4 \pm 0.4) \times 10^{6} \, \mathrm{M}_\odot$ central dark mass \citep{gillessen:2009:mso}, while the highly eccentric, low-periapsis orbit of young star S2 requires a central-mass density $> 10^{13} \, M_\odot \, \mathrm{pc}^{-3}$ \citep{maoz:1998:dc}; a density $> 10^{13} \, M_\odot \, \mathrm{pc}^{-3}$ is also inferred from the compactness of the radio source \citep{genzel:2010RvMP...82.3121G}. These limits provide compelling evidence that the dark point-mass at SgrA$^*$ is an MBH \citep{maoz:1998:dc,genzel:2000MNRAS.317..348G,genzel:2006Natur.442..786G}.

Unfortunately, the nearest large external galaxy is 100 times farther from Earth than SgrA$^*$, and the nearest quasar is 100,000 times farther, so probing other galactic centers is prohibitive. It will however become possible with eLISA.
This is because MBHs are surrounded by a variety of stellar populations, including compact stellar remnants (stellar BHs, NSs, and WDs) that can reach very relativistic orbits around the MBH without being tidally disrupted \cite{amaro-soane:2007:tr}. The compact stars may plunge directly into the event horizon of the MBH; or they may spiral in gradually while emitting GWs. These latter systems, known as \emph{extreme-mass ratio inspirals} (EMRIs), will produce signals detectable by eLISA for MBH masses of $10^4\mbox{--}10^7\, M_\odot$. Stellar-mass BHs should be concentrated in cusps near MBHs \citep{sigurdsson:1997:csm,miralda-escoude:2000:bhgc,freitag:2006JPhCS..54..252F,freitag:2006:srg,hopman:2006:rrn} and generate stronger GWs thanks to their relatively larger mass, so they will provide most detections.

EMRIs are produced when compact stars in the inner 0.01 pc of galactic nuclei are repeatedly scattered by other stars into highly eccentric orbits where gravitational radiation takes over their evolution \cite{amaro-soane:2007:tr}; \emph{resonant relaxation} caused by long-term torques between orbits increases the rate of orbit diffusion \citep{hopman:2006:ems,guerkan:2007:rr}, although relativistic precession can hinder this mechanism \citep{merritt:2011PhRvD..84d4024M}. EMRIs can also be made from the tidal disruption of binaries that pass close to the MBH \citep{miller:2005:bes}, possibly ejecting the hypervelocity stars observed in our Galaxy (see, e.g., \cite{brown:2009:asd}); and from massive-star formation and rapid evolution in the MBH's accretion disk \citep{levin:2007:ssmbh}. Different mechanisms will lead to different EMRI eccentricities and inclinations, evident in the GW signal \citep{miller:2005:bes}.

The detection of even a few EMRIs will provide a completely new probe of dense stellar systems, characterizing the mechanisms that shape stellar dynamics in the galactic nuclei, and recovering information about the MBH, the compact object, and the EMRI orbit with unprecedented precision \citep{amaro-soane:2007:tr}. 
Especially coveted prizes will be accurate masses for $10^5\mbox{--}10^7 \, M_\odot$ MBHs in small, non-active galaxies, which will shed light on galaxy--MBH correlations at the low-mass end; MBH spins, which will illuminate the mechanism of MBH growth by mergers and accretion (see Sec.\ \ref{sec:massiveblackholes}); as well as stellar-BH masses, which will provide insight on stellar formation in the extreme conditions of dense galactic nuclei.
The key to measurement precision is the fact that the compact object behaves as a test particle in the background MBH geometry over hundreds of thousands of relativistic orbits in a year; the resulting GW radiation encodes the details of both the geometry and the orbit \citep{ryan:1995:gwi,ryan:1997,barack:2007,finn:2000:gwc}.

To assess the eLISA science performance on EMRIs, we model their very complicated signals \cite{2006CQGra..23S.769D} using the Barack--Cutler (BC) phenomenological waveforms \citep{barack:2004:lcs}, which are not sufficiently accurate for detection, but capture the character and complexity of EMRI waveforms. We complement this analysis with more realistic \emph{Teukolsky-based} (TB) waveforms obtained by solving the perturbative equations for the BH geometry in the presence of the inspiraling body \citep{teukolsky:1973:rbh}; these have been tabulated for circular--equatorial orbits and for some values of MBH spin \citep{finn:2000:gwc,gair:2009CQGra..26i4034G}. 

To evaluate expected EMRI detection horizons and detection rates, we perform a Monte Carlo over 500,000 realizations of the source parameters, taking MBH rest mass in $[10^{4},5\times 10^6]\, M_\odot$ with a uniform $\log M_\bullet$ distribution; MBH spin uniformly in $[0,0.95]$; compact-body mass of $10 \, M_\odot$, representative of a stellar-mass BH; orbit eccentricity before the final plunge uniformly in $[0.05,0.4]$; and all orbital angles and phases with the appropriate uniform distributions on the circle or sphere, with an equal number of prograde and retrograde orbits. We take the poorly known EMRI formation rate to scale with MBH mass as $400 \, \mathrm{Gyr}^{-1} (M_\bullet/3 \times 10^{6} \, M_\odot)^{-0.19}$ \cite{hopman:2009:emri,preto:2010:ApJ...708L..42P,amaro-seoane:2011CQGra..28i4017A}, and we distribute systems uniformly in comoving volume. Our assumptions are consistent with the MBH mass function derived from the observed galaxy luminosity function using the $M_\bullet\mbox{--}\sigma$ relation, and excluding Sc-Sd galaxies \citep{Aller:2002,gair:2004:ere,gair:2009CQGra..26i4034G}. We further assume an observation time of two years, consider EMRIs in the last five years of their orbit \citep{gair:2009CQGra..26i4034G}, and require a detection $\mathrm{SNR} = 20$ \cite{cornish:2011CQGra..28i4016C,gair:2008:cmh,babak:2010:mldc}.

\begin{figure}
\flushright
\includegraphics[width=0.8\textwidth]{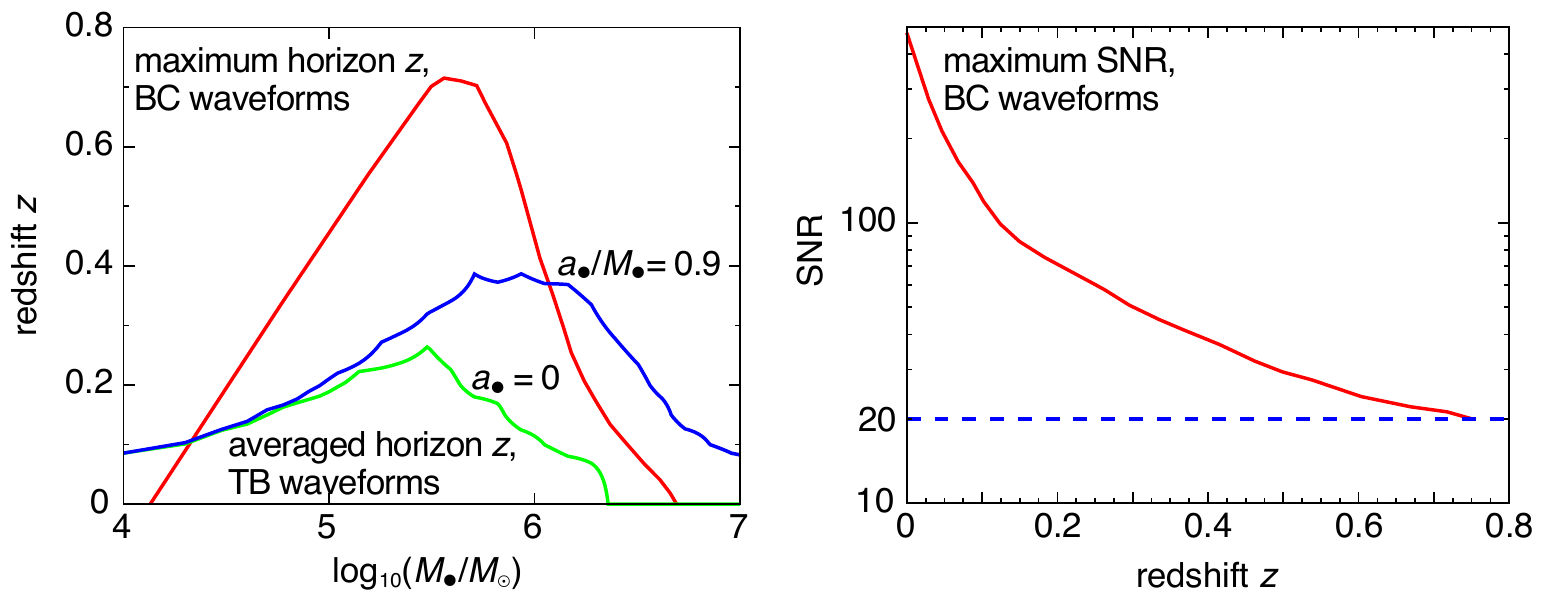}
\caption{\textbf{Left}: maximum detection horizon redshift vs.\ MBH rest mass, BC EMRI waveforms (red curve); averaged horizon redshift vs.\ MBH rest mass, TB EMRI waveforms with $a_\bullet/M_\bullet = 0$ and $0.9$. Assumptions are given in the main text; the maximum is computed as the highest $z$ with $\mathrm{SNR} > 20$ in a given mass bin. \textbf{Right}: maximum EMRI SNR vs.\ redshift, BC waveforms.\label{fig:emrihorizon}}
\end{figure}

The left panel of Fig.\ \ref{fig:emrihorizon} shows the resulting \emph{maximum} horizon redshift for BC waveforms, as a function of MBH rest mass---that is, it shows the $z$ at which an optimally oriented source with the most favorable MBH and orbit parameters (as found in the Monte Carlo) achieves the detection SNR. Thus, EMRIs in the eLISA range will be detectable as far $z = 0.7$. By contrast, EM observations of $10^4\mbox{--}10^6 \, M_\odot$ MBHs are possible in the local Universe out to $z \simeq 0.1$. The right panel plots the distribution of SNRs as a function of $z$, which shows that nearby EMRIs in the local Universe will yield SNRs of many tens.

For comparison, the left panel of Fig.\ \ref{fig:emrihorizon} shows also the horizons computed with sky- and orientation-averaged SNRs, using TB waveforms from circular--equatorial orbits with MBH spins $a_\bullet/M_\bullet = 0$ and $0.9$. The difference between the BC and TB curves is consistent with the effects of sky-averaging: SNRs for optimally oriented systems are expected to be 2.5 times higher than averaged SNRs. The $a_\bullet/M_\bullet = 0.9$ systems are favored because high MBH spin allows for orbits closer to the event horizon and higher GW frequencies, which shifts the peak eLISA sensitivity to higher masses.

The resulting number of expected eLISA detections over two years is $\sim 50$, as evaluated with the BC-waveform Monte Carlo, and $\sim 30/35/55$ (for $a_\bullet/M_\bullet = 0/0.5/0.9$), as evaluated with TB-waveform sky-averaged horizons. The higher TB event rate is explained by the inclusion of eccentric systems, which radiate more energy in the eLISA band, and it should be more reliable because of the broad sampling of source parameters. Remember however that EMRI rates are highly uncertain \cite{amaro-soane:2007:tr,hopman:2009:emri,preto:2010:ApJ...708L..42P,merritt:2011PhRvD..84d4024M}. Even with as few as 10 events, the slope of the MBH mass function in the $10^4\mbox{--}10^6 \, M_\odot$ range can be determined to 0.3, the current level of observational uncertainty \cite{gair:2010:emri}.
\begin{figure}
\flushright
\includegraphics[width=0.8\textwidth]{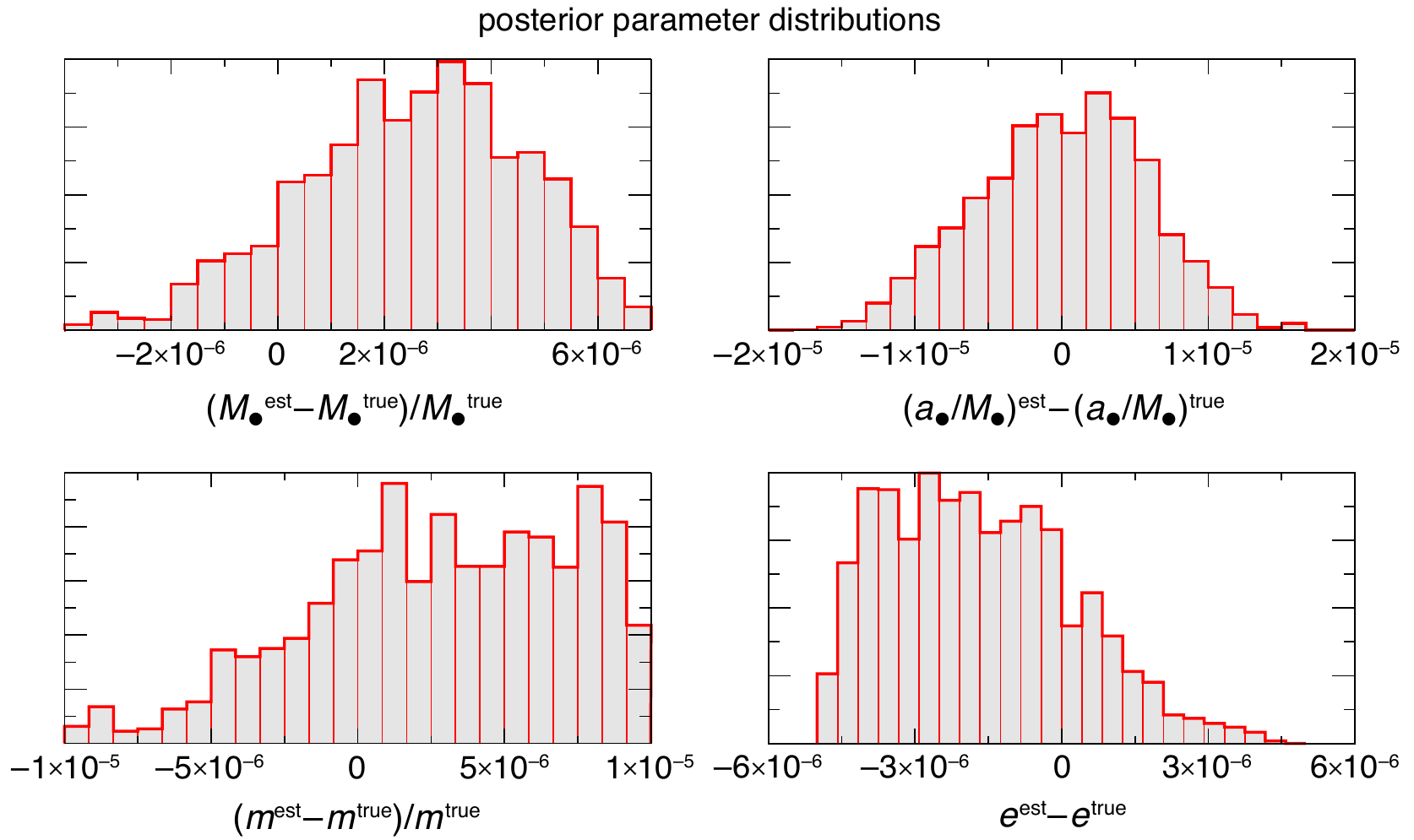}
\caption{Posterior probability plot for source parameters (MBH rest mass $M_\bullet$, MBH spin $a_\bullet$, compact-body mass $m$, and orbit eccentricity at plunge $e$), in the $\mathrm{SNR} = 25$ detection of a $10 + 10^6 \, M_\odot$ EMRI at $z = 0.55$, with $a_\bullet/M_\bullet = 0.7$ and $e_\mathrm{plunge} = 0.25$.
\label{fig:emrimcmc}}
\end{figure}

Because EMRI waveforms are such complex and sensitive functions of the source parameters, these will be estimated accurately whenever an EMRI is detected \citep{cornish:2011CQGra..28i4016C,gair:2008:cmh,babak:2010:mldc}. In particular, we expect to measure the MBH mass and spin, as well as the compact-body mass and eccentricity to better than a part in $10^3$ \citep{barack:2004:lcs}. As an example, Fig.\ \ref{fig:emrimcmc} shows the posterior distributions of the best-determined parameters for a $z = 0.55$ source detected by eLISA with $\mathrm{SNR} = 25$, as computed with the Markov Chain Monte Carlo algorithm of \citep{cornish:2006:mes}; for this source, the luminosity distance $D_L$ would be determined to 1\%, and the sky location to 0.2 $\mathrm{deg}^2$. Even with relatively low SNR, parameter-estimation accuracy is excellent. In general, we find that the eLISA and LISA parameter-estimation performance is very similar for EMRIs detected with the same SNR (but of course different distances), so the reader can refer to treatments for LISA in the literature \citep{barack:2004:lcs,huerta:2009PhRvD..79h4021H,porter:2009GWN.....1....4P,babak:2010:mldc}.

\section{Precision measurements of strong gravity}
\label{sec:gravity}

Einstein's theory of gravity, general relativity (GR), has been tested rigorously in the Solar system and in binary pulsars \citep{will:2006:gre,lorimer:2008:bmp}; these tests, however, probe only the weak-field regime where the characteristic perturbative parameter $\epsilon = v^2/c^2 \sim G M / (R c^2)$ is very small, $\sim 10^{-6}\mbox{--}10^{-8}$ (here $v$ is the velocity of gravitating bodies, $M$ their mass, and $R$ their separation). By contrast, eLISA's GW observations of coalescing MBHs (Sec.\ \ref{sec:massiveblackholes}) and of EMRIs (Sec.\ \ref{sec:emris}) will allow us to confront GR with precision measurements of its dynamical, strong-field regime, and to verify that astrophysical BHs are really the Kerr mathematical solutions predicted by GR.

Before considering the GR tests possible with each of these sources, we note that, by the second half of this decade, second-generation ground-based detectors are expected to routinely observe the coalescences of stellar-mass BHs and (possibly) of asymmetric systems such as a NS inspiraling into a $100 \, M_\odot$ BH. However, they will do so with 10--100 times lower SNRs than eLISA (for the brightest sources), and for up to 1,000 times fewer GW cycles; thus, eLISA will test our understanding of gravity in the most extreme conditions with a precision that is two orders of magnitude better than that achievable from the ground. 
(Although most of the references cited in the rest of this section were developed for LISA, their broad conclusions are applicable to sources detected with comparable SNRs by eLISA.)

All three phases of MBH coalescence offer opportunities for precision measurements.
The year-long \textbf{inspiral signals} can be examined for evidence of a massive graviton, resulting in a frequency-dependent phase shift of the waveform, improving current Solar-system bounds \citep{2011PhRvD..84j1501B,huwyler:2011arXiv1108.1826H};
they can yield stringent constraints on other theories with deviations from GR parametrized by a set of global parameters, such as massless and massive Brans-Dicke theories \citep{berti:2005:esb,2012PhRvD..85f4041A}, theories with an evolving gravitational constant \citep{yunes:2010PhRvD..81f4018Y}, Lorentz-violating modifications of GR \citep{2012PhRvD..85b4041M};
last, various authors have considered testing inspiral waves for hypothetical, generic modifications of their amplitude and phasing \citep{arun:2006:pns,yunes:2009PhRvD..80l2003Y,cornish:2011PhRvD..84f2003C,2012PhRvD..85h2003L}.

The \textbf{merger} of comparable-mass MBH binaries produces an enormously powerful GW burst, which eLISA will measure with SNR as high as a few hundred, even at cosmological distances. The MBH masses and spin can be determined with high accuracy from the inspiral waveform; given these physical parameters, numerical relativity can predict the shape of the merger waveform, as well as the mass and spin of the final remnant MBH \citep{rezzolla:2008ApJ...674L..29R}, and these can be compared directly with observations, providing an ideal test of pure GR in a highly dynamical, strong-field regime.

The frequencies and damping times of the quasinormal modes (QNMs) in the final \textbf{ringdown} \cite{berti:2009CQGra..26p3001B} are completely determined by the mass and the spin of the remnant, and therefore can be used to measure them \cite{berti:2006:gws,berti:2007PhRvD..76j4044B}, while their relative amplitudes hold information about the pre-merger binary \citep{2012PhRvD..85b4018K}, again providing a check of consistency between GR predictions for the phases of coalescence. Furthermore, the measurement of at least two QNMs \citep{berti:2007PhRvD..76j4044B} will test the Kerr-ness of the MBH \citep{dreyer:2004:bhs} against exotic proposals such as boson stars and gravastars \citep{yoshida:1994PhRvD..50.6235Y,berti:2006:boson,chirenti:2007CQGra..24.4191C,pani:2009PhRvD..80l4047P}. Modifications of GR that lead to different emission would also be apparent \citep{barausse:2008xv,pani:2009PhRvD..79h4031P}.

\textbf{EMRIs} are expected to be very clean astrophysical systems, except
perhaps in few systems with strong interactions with the accretion disk \citep{barausse:2007PhRvD..75f4026B,barausse:2008PhRvD..77j4027B,kocsis:2011PhRvD..84b4032K}, or with perturbations due to a second nearby MBH or star \citep{yunes:2011PhRvD..83d4030Y,2012ApJ...744L..20A}. Over day-long timescales, EMRI orbits are essentially geodesics of the background geometry; on longer timescales, the loss of energy and angular momentum to GWs causes a slow change of the geodesic parameters. In the last few years of their evolution, as observed by eLISA, EMRI orbits are highly relativistic ($R < 10\, R_\bullet$) and display extreme forms of periastron and orbital plane precession.
Indeed, EMRI GWs encode all the mass and current multipoles of the MBH \citep{ryan:1995:gwi,drasco:2004:rbh}, which for a Kerr BH are uniquely determined by the mass and spin alone (another manifestation of the ``no-hair'' theorem). 
For EMRIs with $\mathrm{SNR} = 30$, eLISA will measure mass and spin to a part in $10^3\mbox{--}10^4$, and the mass quadrupole moment $M_2$ to a part in $10^2\mbox{--}10^4$, thus \textbf{testing the no-hair theorem} directly \cite{barack:2007}. See \citep{sopuerta:2010:emri,babak:2011CQGra..28k4001B} for reviews of different ways to test the nature of astrophysical BHs.

Other tests of the Kerr-ness of the central massive object have been proposed: for a boson star, the EMRI signal would not shut off after the last stable orbit \citep{kesden:2005:gws}; for a gravastar, QNMs could be excited resonantly \citep{pani:2009PhRvD..80l4047P}; for certain non-Kerr axisymmetric geometries, orbits could become ergodic or experience resonances \citep{gair:2008:pbh,lukes-gerakopoulos:2010PhRvD..81l4005L}; for ``bumpy'' BHs, orbits would again carry distinctive signatures \citep{ryan:1995:gwi,collins:2004:tfm,glampedakis:2006:msl,vigeland:2011PhRvD..83j4027V}. Modifications in EMRI GWs would also arise if the \textbf{true theory of gravity is in fact different from GR}, as are dynamical Chern-Simons theory \citep{sopuerta:2009PhRvD..80f4006S,pani:2011PhRvD..83j4048P}, scalar--tensor theories (with observable effects in NS--BH systems where the NS carries scalar charge \citep{berti:2005:esb,yagi:2010PhRvD..81f4008Y}), Randall--Sundrum-inspired braneworld models \citep{mcwilliams:2010PhRvL.104n1601M,yagi:2011PhRvD..83h4036Y}, theories with axions that give rise to ``floating orbits'' \citep{Cardoso:2011xi,2012PhRvD..85j2003Y}, as well as generic, phenomenologically parametrized theories \citep{gair:2011PhRvD..84f4016G}.

\section{Cosmology and new physics from the early Universe}
\label{sec:cosmo}

GWs produced after the Big Bang form a fossil radiation: expansion prevents them from reaching thermal equilibrium with the other components because of the weakness of the gravitational interaction. Thus, relic GWs carry information about the first instants of the Universe. If their wavelength is set by the apparent horizon size
$c/H_* = c (a/\dot{a})_*$
at the time of production, when the temperature of the Universe is $T_*$, the 
redshifted frequency is
\begin{equation}
    f \approx 10^{-4}\,\mathrm{Hz} \,
    \sqrt{H_* \times \frac{1 \, \mathrm{mm}}{c}} \approx 10^{-4}\,\mathrm{Hz}
  \left(\frac{k_B T_*}{1 \, \mathrm{TeV}}\right),
\end{equation}
so the eLISA frequency band corresponds to the horizon at and beyond the \emph{Terascale frontier} of fundamental physics.  This allows eLISA to probe bulk motions at times about $3 \times 10^{-18}\mbox{--}3 \times 10^{-10}$ s after the Big Bang, a period not directly accessible with any other technique.  Taking a typical broad spectrum into account, eLISA has the sensitivity to detect cosmological backgrounds caused by \emph{new physics} at energies $\sim 0.1\mbox{--}1000\,\mathrm{TeV}$, if more than a (modest) fraction $\sim 10^{-5}$ of the energy density is converted to GWs at the time of production.

\begin{figure}
  \flushright
  \includegraphics[width=0.7\textwidth]{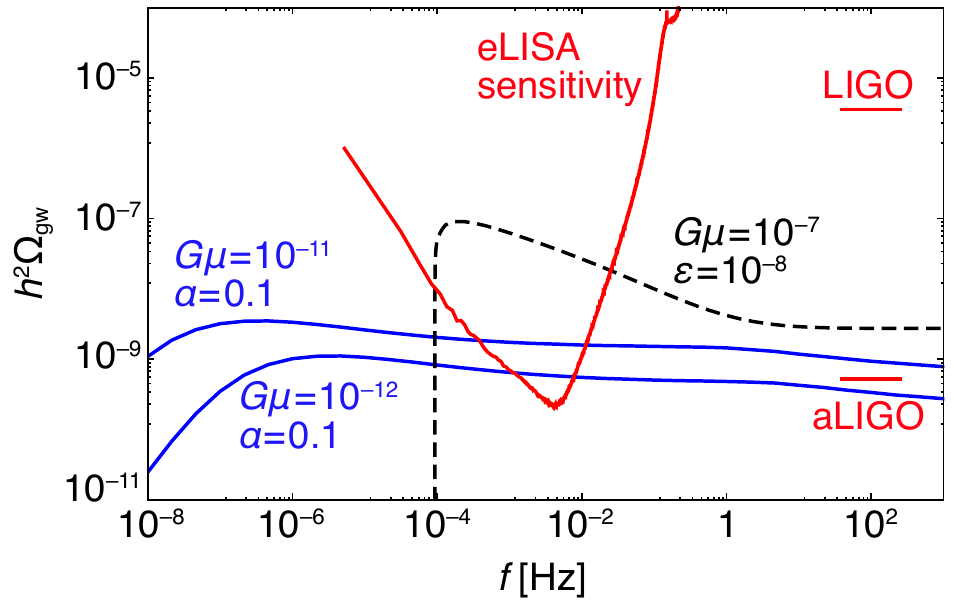}
  \caption{(From \cite{2012JCAP...06..027B}.) Spectra of stochastic backgrounds from cosmic strings for large loops (with horizon size $\alpha = 0.1$, solid lines), for two values of the string tension $G\mu/c^4$ spanning a range of scenarios motivated by braneworld inflation; and for small loops (with size $\alpha = 50 \epsilon G \mu$, dashed line). The cosmic-string spectrum is distinguishably different from that of first-order phase transitions or any other predicted source: it has nearly constant energy per logarithmic frequency interval over many decades at high frequencies, and falls off after a peak at low frequencies, since  large string loops are rare and radiate slowly. Cosmic strings may also produce distinctive bursts, produced by a sharply bent bits of string moving at nearly the speed of light \citep{damour:2005:grc,siemens:2006:gwb,binetruy:2010PhRvD..82l6007B,bohe:2011PhRvD..84f5016B}.
\label{fig:strings}}
\end{figure}

Various sources of \textbf{cosmological GW backgrounds} are presented in detail in \cite{2012JCAP...06..027B}. They include first-order phase transitions, resulting in bubble nucleation and growth, and subsequent bubble collisions and turbulence \citep{witten:1984:csp,hogan:1986:grc,Kamionkowski:1993fg,Huber:2008hg,Caprini:2009yp}; the dynamics of stabilization for the extra dimensions required by superstring theory \citep{hogan:2000:gwm,randall:2006:gww}, which may also appear as non-Newtonian gravity in laboratory experiments at the sub-mm scale;
networks of cosmic (super-)strings \citep{copeland:2004JHEP...06..013C,1994csot.book.....V}, which continuously produce loops that decay into GWs (see Fig.\ \ref{fig:strings});
the transition between inflation and the hot Big Bang in the process of preheating \citep{khlebnikov:1997:rgw,easther:2006:sgw,GarciaBellido:2007dg,Dufaux:2007pt,Dufaux:2008dn}; and the amplification of quantum vacuum fluctuations in some unconventional versions of inflation \cite{brustein:1995:rgw,buonanno:2003:tasi,buonanno:1997:srg}. Although the two-arm eLISA does not provide a Sagnac observable \citep{hogan:2001:esg} to calibrate instrument noise against possible GW backgrounds, the clear spectral dependence predicted for some of these phenomena provides an observational handle, as long as the background lies above the eLISA sensitivity curve.%

As discussed in Sec.\ \ref{sec:massiveblackholes}, observations of GWs from MBH binaries probe the assembly of cosmic structures. In addition, binaries can serve as \emph{standard sirens} to \textbf{measure cosmological parameters} \citep{schutz:1986:hgw,holz:2005:ugw} because, as discussed around Eq.\ \eqref{eq:fders}, measuring the amplitude and frequency evolution of a binary signal yields the absolute luminosity distance to the source. However, binary GWs cannot provide the source's redshift unless the other source parameters are known independently (because the rest mass of the binary is the only length/time scale in the waveform, the frequency evolution of a redshifted signal is indistinguishable from the signal from a heavier binary). The optical redshift of the host galaxy can be obtained if an EM counterpart to MBH coalescence is observed (see, e.g., \cite{armitage:2002ApJ...567L...9A,milosavljevic:2005:amb,phinney:2009astro2010S.235P}, and \citep{schnittman:2011CQGra..28i4021S} for a recent review).

While there are many uncertainties in the nature and strength of such counterparts, some may be observable in the local Universe. At $z < 1$, we expect that eLISA MBH-inspiral measurements could provide sky locations to better than 400 $\mathrm{deg}^2$ for 50\% of sources, and to 10 $\mathrm{deg}^2$ for 11\%. (The inclusion of merger and ringdown in the analysis should further improve these numbers.) Such large areas will be covered frequently and deeply by optical and radio surveys such as LSST \citep{lsst:2009arXiv0912.0201L} and the VAST project \citep{johnston:2007PASA...24..174J}, identifying sufficiently distinctive transients. The accurate knowledge of the counterpart's redshift and position would then improve the uncertainty of GW-determined parameters, with $D_L$ known to 1\% for 60\% of sources, and 5\% for 87\%. Such precise luminosity distance--redshift measurements will be complementary to other cosmographical campaigns \citep{riess:1998ApJ...504..935R,perlmutter:1999AIPC..478..129P}, and will improve the estimation of cosmological parameters. Even without counterparts, one may proceed by considering all possible hosts in a distance--position error box, and enforcing consistency between multiple GW events \citep{petiteau:2011ApJ...732...82P}; this should be possible for MBH binaries (and EMRIs \citep{macleod:2008:phc}) in the local Universe, yielding the Hubble constant to a few percent.

\section{Conclusions}
\label{sec:conclusions}

While LISA was always meant to be the definitive mission in its frequency band, eLISA is being designed to provide the maximum science within a cost cap. Nevertheless, as described above, eLISA will achieve a great part of the LISA science goals. It will represent the culmination of twenty years of exciting, painstaking work, pioneering the new science of observational low-frequency GW astronomy. It will truly begin to unveil the hidden, distant Universe. May it fly soon, and safe.

\ack This research was supported by the Deutsches Zentrum f\"ur L\"uft- und Raumfahrt and by the Transregio 7 ``Gravitational Wave Astronomy'' financed by the Deutsche Forschungsgemeinschaft DFG (German Research Foundation). EB was supported by NSF Grant PHY-0900735 and by NSF CAREER Grant PHY-1055103. AK was supported by the Swiss National Science Foundation. TBL was supported by NASA Grant 08-ATFP08-0126. RNL was supported by an appointment to the NASA Postdoctoral Program at the Goddard Space Flight Center, administered by Oak Ridge Associated Universities through a contract with NASA. MV performed this work at the Jet Propulsion Laboratory, California Institute of Technology, under contract with the National Aeronautics and Space Administration. Copyright 2012.

\section*{References}

\bibliography{ngo_x}

\providecommand{\newblock}{}
\begin{thebibliography}{100}
\expandafter\ifx\csname url\endcsname\relax
  \def\url#1{{\tt #1}}\fi
\expandafter\ifx\csname urlprefix\endcsname\relax\def\urlprefix{URL }\fi
\providecommand{\eprint}[2][]{\url{#2}}

\bibitem{lisasciencecase}
Prince T~A {\em et~al.\/} 2009 {LISA: Probing the Universe with Gravitational
  Waves} \url{list.caltech.edu/mission_documents}

\bibitem{Jennrich:2009p1398}
Jennrich O 2009 {\em Class. Quantum Grav.\/} {\bf 26} 153001

\bibitem{lisaads}
{SAO/NASA Astrophysics Data System} 2012 papers mentioning {LISA} in the
  abstract \url{tinyurl.com/lisa-ads}

\bibitem{national2010New}
{US National Research Council} 2010 {New Worlds, New Horizons in Astronomy and
  Astrophysics} \url{http://www.nap.edu/openbook.php?record_id=12951}

\bibitem{2012arXiv1201.3621A}
{Amaro-Seoane} P {\em et~al.\/} 2012  ArXiv:1201.3621

\bibitem{petiteau:2008PhRvD..77b3002P}
{Petiteau} A {\em et~al.\/} 2008 {\em Phys.~Rev.~D\/} {\bf 77} 023002

\bibitem{2009CQGra..26i4030N}
{Nelemans} G 2009 {\em Class. and Quantum Grav.\/} {\bf 26} 094030

\bibitem{marsh:2011CQGra..28i4019M}
{Marsh} T~R 2011 {\em Class.~Quantum~Grav.\/} {\bf 28} 094019

\bibitem{rau:2010ApJ...708..456R}
{Rau} A {\em et~al.\/} 2010 {\em ApJ\/} {\bf 708} 456

\bibitem{levitan:2011ApJ...739...68L}
{Levitan} D {\em et~al.\/} 2011 {\em ApJ\/} {\bf 739} 68

\bibitem{jonker:2011ApJS..194...18J}
{Jonker} P~G {\em et~al.\/} 2011 {\em ApJS\/} {\bf 194} 18

\bibitem{stroeer:2006:lvb}
{Stroeer} A and {Vecchio} A 2006 {\em Class.~Quantum~Grav.\/} {\bf 23} 809

\bibitem{roelofs:2010:se}
{Roelofs} G~H~A {\em et~al.\/} 2010 {\em ApJ\/} {\bf 711} L138

\bibitem{2006ApJ...649..382S}
{Steeghs} D {\em et~al.\/} 2006 {\em ApJ\/} {\bf 649} 382

\bibitem{2011MNRAS.413.3068C}
{Copperwheat} C~M {\em et~al.\/} 2011 {\em MNRAS\/} {\bf 413} 3068

\bibitem{brown:2011ApJ...737L..23B}
{Brown} W~R {\em et~al.\/} 2011 {\em ApJ\/} {\bf 737} L23

\bibitem{PhysRevD.72.042003}
Vallisneri M 2005 {\em Phys. Rev. D\/} {\bf 72}(4) 042003

\bibitem{2012arXiv1201.4613N}
{Nissanke} S, {Vallisneri} M, {Nelemans} G and {Prince} T~A 2012 {\em ApJ\/}
  {in print, arXiv:1201.4613}

\bibitem{yu:2010:gw}
{Yu} S and {Jeffery} C~S 2010 {\em A\&A\/} {\bf 521} A85

\bibitem{ruiter:2010ApJ...717.1006R}
{Ruiter} A~J {\em et~al.\/} 2010 {\em ApJ\/} {\bf 717} 1006

\bibitem{roelofs:2007:sdss}
{Roelofs} G~H~A, {Nelemans} G and {Groot} P~J 2007 {\em MNRAS\/} {\bf 382} 685

\bibitem{pakmor:2010:sltIa}
{Pakmor} R {\em et~al.\/} 2010 {\em Nature\/} {\bf 463} 61

\bibitem{perets:2010Natur.465..322P}
{Perets} H~B {\em et~al.\/} 2010 {\em Nature\/} {\bf 465} 322

\bibitem{waldman:2011ApJ...738...21W}
{Waldman} R {\em et~al.\/} 2011 {\em ApJ\/} {\bf 738} 21

\bibitem{webbink:1984:dwd}
{Webbink} R~F 1984 {\em ApJ\/} {\bf 277} 355

\bibitem{levan:2006:grb}
{Levan} A~J {\em et~al.\/} 2006 {\em MNRAS\/} {\bf 368} L1

\bibitem{taam:2000:cee}
{Taam} R~E and {Sandquist} E~L 2000 {\em ARA\&A\/} {\bf 38} 113

\bibitem{taam:2010NewAR..54...65T}
{Taam} R~E and {Ricker} P~M 2010 {\em New Astro. Rev.\/} {\bf 54} 65

\bibitem{nelemans:2005MNRAS.356..753N}
{Nelemans} G and {Tout} C~A 2005 {\em MNRAS\/} {\bf 356} 753

\bibitem{demarco:2011MNRAS.411.2277D}
{De Marco} O {\em et~al.\/} 2011 {\em MNRAS\/} {\bf 411} 2277

\bibitem{paczynski:1976:secbs}
{Paczynski} B 1976 {Common Envelope Binaries} {\em Structure and Evolution of
  Close Binary Systems\/} ({\em IAU Symposium\/} vol~73) ed {P~Eggleton,
  S~Mitton, \& J~Whelan} p~75

\bibitem{stroeer:2009:stv}
{Stroeer} A and {Nelemans} G 2009 {\em MNRAS\/} {\bf 400} L24

\bibitem{marsh:2004:mtwd}
{Marsh} T~R, {Nelemans} G and {Steeghs} D 2004 {\em MNRAS\/} {\bf 350} 113

\bibitem{dsouza:2006:dmt}
{D'Souza} M~C~R, {Motl} P~M, {Tohline} J~E and {Frank} J 2006 {\em ApJ\/} {\bf
  643} 381

\bibitem{racine:2007:ndts}
{Racine} {\'E}, {Phinney} E~S and {Arras} P 2007 {\em MNRAS\/} {\bf 380} 381

\bibitem{nelemans:2001:ps2}
{Nelemans} G, {Portegies Zwart} S~F, {Verbunt} F and {Yungelson} L~R 2001 {\em
  A\&A\/} {\bf 368} 939

\bibitem{edlund:2005:wdw}
{Edlund} J~A, {Tinto} M, {Kr{\'o}lak} A and {Nelemans} G 2005 {\em
  Phys.~Rev.~D\/} {\bf 71} 122003

\bibitem{alexander:2005:spmbh}
{Alexander} T 2005 {\em Phys.~Rep.\/} {\bf 419} 65

\bibitem{ruiter:2009:hwdb}
{Ruiter} A~J, {Belczynski} K, {Benacquista} M and {Holley-Bockelmann} K 2009
  {\em ApJ\/} {\bf 693} 383

\bibitem{belczynski:2010:dco}
{Belczynski} K, {Benacquista} M and {Bulik} T 2010 {\em ApJ\/} {\bf 725} 816

\bibitem{2001A&A...369..339P}
{Perryman} M~A~C 2001 {\em Astronomy and Astrophysics\/} {\bf 369} 339

\bibitem{Salpeter:1964}
{Salpeter} E~E 1964 {\em ApJ\/} {\bf 140} 796

\bibitem{zeldovich:1964:rgw}
{Zel'dovich} Y~B and {Novikov} I~D 1964 {\em Soviet Physics Doklady\/} {\bf 9}
  246

\bibitem{krolik:1999:agn}
{Krolik, J~H} (ed) 1999 {\em {Active galactic nuclei: from the central black
  hole to the galactic environment}\/} (Princeton, N.J.: Princeton University
  Press)

\bibitem{gultekin:2009:ms}
{G{\"u}ltekin} K {\em et~al.\/} 2009 {\em ApJ\/} {\bf 698} 198

\bibitem{dimatteo:2005:eiq}
{Di Matteo} T, {Springel} V and {Hernquist} L 2005 {\em Nature\/} {\bf 433} 604

\bibitem{hopkins:2006:umm}
{Hopkins} P~F {\em et~al.\/} 2006 {\em ApJS\/} {\bf 163} 1

\bibitem{croton:2006MNRAS.365...11C}
{Croton} D~J {\em et~al.\/} 2006 {\em MNRAS\/} {\bf 365} 11

\bibitem{white:1978MNRAS.183..341W}
{White} S~D~M and {Rees} M~J 1978 {\em MNRAS\/} {\bf 183} 341

\bibitem{haiman:1998ApJ...503..505H}
{Haiman} Z and {Loeb} A 1998 {\em ApJ\/} {\bf 503} 505

\bibitem{haehnelt:1998MNRAS.300..817H}
{Haehnelt} M~G, {Natarajan} P and {Rees} M~J 1998 {\em MNRAS\/} {\bf 300} 817

\bibitem{wyithe:2002ApJ...581..886W}
{Wyithe} J~S~B and {Loeb} A 2002 {\em ApJ\/} {\bf 581} 886

\bibitem{volonteri:2003:amh}
{Volonteri} M, {Haardt} F and {Madau} P 2003 {\em ApJ\/} {\bf 582} 559

\bibitem{madau:2001:mbhIII}
{Madau} P and {Rees} M~J 2001 {\em ApJ\/} {\bf 551} L27

\bibitem{haehnelt:1993MNRAS.263..168H}
{Haehnelt} M~G and {Rees} M~J 1993 {\em MNRAS\/} {\bf 263} 168

\bibitem{loeb:1994:cbg}
{Loeb} A and {Rasio} F~A 1994 {\em ApJ\/} {\bf 432} 52

\bibitem{devecchi:2009ApJ...694..302D}
{Devecchi} B and {Volonteri} M 2009 {\em ApJ\/} {\bf 694} 302

\bibitem{mayer:2010Natur.466.1082M}
{Mayer} L, {Kazantzidis} S, {Escala} A and {Callegari} S 2010 {\em Nature\/}
  {\bf 466} 1082

\bibitem{volonteri:2010A&ARv..18..279V}
{Volonteri} M 2010 {\em A\&A~Rev.\/} {\bf 18} 279

\bibitem{2012AdAst2012E..12S}
{Sesana} A 2012 {\em Adv. Astron.\/} {\bf 2012} 805402

\bibitem{marconi:2004:lsmbh}
{Marconi} A {\em et~al.\/} 2004 {\em MNRAS\/} {\bf 351} 169

\bibitem{yu:2002:oc}
{Yu} Q and {Tremaine} S 2002 {\em MNRAS\/} {\bf 335} 965

\bibitem{soltan:1982:moq}
{Soltan} A 1982 {\em MNRAS\/} {\bf 200} 115

\bibitem{volonteri:2007:bhs}
{Volonteri} M, {Sikora} M and {Lasota} J 2007 {\em ApJ\/} {\bf 667} 704

\bibitem{shakura:1973A&A....24..337S}
{Shakura} N~I and {Sunyaev} R~A 1973 {\em A\&A\/} {\bf 24} 337

\bibitem{thorne:1974:dabh}
{Thorne} K~S 1974 {\em ApJ\/} {\bf 191} 507

\bibitem{gammie:2004:bhse}
{Gammie} C~F, {Shapiro} S~L and {McKinney} J~C 2004 {\em ApJ\/} {\bf 602} 312

\bibitem{king:2006MNRAS.373L..90K}
{King} A~R and {Pringle} J~E 2006 {\em MNRAS\/} {\bf 373} L90

\bibitem{berti:2008:cbhse}
{Berti} E and {Volonteri} M 2008 {\em ApJ\/} {\bf 684} 822

\bibitem{begelman:1980Natur.287..307B}
{Begelman} M~C, {Blandford} R~D and {Rees} M~J 1980 {\em Nature\/} {\bf 287}
  307

\bibitem{chandrasekhar:1943:dfI}
{Chandrasekhar} S 1943 {\em ApJ\/} {\bf 97} 255

\bibitem{ostriker:1999ApJ...513..252O}
{Ostriker} E~C 1999 {\em ApJ\/} {\bf 513} 252

\bibitem{colpi:1999ApJ...525..720C}
{Colpi} M, {Mayer} L and {Governato} F 1999 {\em ApJ\/} {\bf 525} 720

\bibitem{mayer:2007Sci...316.1874M}
{Mayer} L {\em et~al.\/} 2007 {\em Science\/} {\bf 316} 1874

\bibitem{quinlan:1996:dembhb}
{Quinlan} G~D 1996 {\em NewA\/} {\bf 1} 35

\bibitem{khan:2011ApJ...732...89K}
{Khan} F~M, {Just} A and {Merritt} D 2011 {\em ApJ\/} {\bf 732} 89

\bibitem{preto:2011ApJ...732L..26P}
{Preto} M, {Berentzen} I, {Berczik} P and {Spurzem} R 2011 {\em ApJ\/} {\bf
  732} L26

\bibitem{escala:2004:rgm}
{Escala} A, {Larson} R~B, {Coppi} P~S and {Mardones} D 2004 {\em ApJ\/} {\bf
  607} 765

\bibitem{dotti:2007:smbh}
{Dotti} M, {Colpi} M, {Haardt} F and {Mayer} L 2007 {\em MNRAS\/} {\bf 379} 956

\bibitem{cuadra:2009MNRAS.393.1423C}
{Cuadra} J, {Armitage} P~J, {Alexander} R~D and {Begelman} M~C 2009 {\em
  MNRAS\/} {\bf 393} 1423

\bibitem{2011arXiv1107.2819S}
{Sperhake} U, {Berti} E and {Cardoso} V 2011  ArXiv:1107.2819

\bibitem{2011PhRvL.107w1102L}
{Lousto} C~O and {Zlochower} Y 2011 {\em Phys. Rev. Lett.\/} {\bf 107} 231102

\bibitem{haehnelt:1994:lfg}
{Haehnelt} M~G 1994 {\em MNRAS\/} {\bf 269} 199

\bibitem{wyithe:2003:lfg}
Wyithe J~S~B and Loeb A 2003 {\em ApJ\/} {\bf 590} 691

\bibitem{sesana:2004:lfg}
{Sesana} A, {Haardt} F, {Madau} P and {Volonteri} M 2004 {\em ApJ\/} {\bf 611}
  623

\bibitem{enoki:2004:gws}
{Enoki} M, {Inoue} K~T, {Nagashima} M and {Sugiyama} N 2004 {\em ApJ\/} {\bf
  615} 19

\bibitem{sesana:2005:gws}
{Sesana} A, {Haardt} F, {Madau} P and {Volonteri} M 2005 {\em ApJ\/} {\bf 623}
  23

\bibitem{rhook:2005:rer}
{Rhook} K~J and {Wyithe} J~S~B 2005 {\em MNRAS\/} {\bf 361} 1145

\bibitem{koushiappas:2006:tms}
{Koushiappas} S~M and {Zentner} A~R 2006 {\em ApJ\/} {\bf 639} 7

\bibitem{sesana:2007:imprint}
{Sesana} A, {Volonteri} M and {Haardt} F 2007 {\em MNRAS\/} {\bf 377} 1711

\bibitem{flanagan:1998:mgwa}
{Flanagan} {\'E}~{\'E} and {Hughes} S~A 1998 {\em Phys.~Rev.~D\/} {\bf 57} 4535

\bibitem{santamaria:2010PhRvD..82f4016S}
{Santamar{\'{\i}}a} L {\em et~al.\/} 2010 {\em Phys.~Rev.~D\/} {\bf 82} 064016

\bibitem{blanchet:2006:grp}
Blanchet L 2006 {\em Living Reviews in Relativity\/} {\bf 9} 3

\bibitem{arun:2009:petf}
{Arun} K~G {\em et~al.\/} 2009 {\em Class.~Quantum~Grav.\/} {\bf 26} 094027

\bibitem{dotti:2010MNRAS.402..682D}
{Dotti} M {\em et~al.\/} 2010 {\em MNRAS\/} {\bf 402} 682

\bibitem{McClintock:2011zq}
McClintock J~E {\em et~al.\/} 2011 {\em Class. Quantum Grav.\/} {\bf 28} 114009

\bibitem{sesana:2011:rmbh}
{Sesana} A, {Gair} J, {Berti} E and {Volonteri} M 2011 {\em Phys.~Rev.~D\/}
  {\bf 83} 044036

\bibitem{berti:2006:gws}
{Berti} E, {Cardoso} V and {Will} C~M 2006 {\em Phys.~Rev.~D\/} {\bf 73} 064030

\bibitem{portegieszwart:2000ApJ...528L..17P}
{Portegies Zwart} S~F and {McMillan} S~L~W 2000 {\em ApJ\/} {\bf 528} L17

\bibitem{guerkan:2004ApJ...604..632G}
{G{\"u}rkan} M~A, {Freitag} M and {Rasio} F~A 2004 {\em ApJ\/} {\bf 604} 632

\bibitem{portegieszwart:2004Natur.428..724P}
{Portegies Zwart} S~F {\em et~al.\/} 2004 {\em Nature\/} {\bf 428} 724

\bibitem{freitag:2006JPhCS..54..252F}
{Freitag} M, {Amaro-Seoane} P and {Kalogera} V 2006 {\em JPCS\/} {\bf 54} 252

\bibitem{freitag:2006MNRAS.368..121F}
{Freitag} M, {Rasio} F~A and {Baumgardt} H 2006 {\em MNRAS\/} {\bf 368} 121

\bibitem{guerkan:2006ApJ...640L..39G}
{G{\"u}rkan} M~A, {Fregeau} J~M and {Rasio} F~A 2006 {\em ApJ\/} {\bf 640} L39

\bibitem{amaro-seoane:2006ApJ...653L..53A}
{Amaro-Seoane} P and {Freitag} M 2006 {\em ApJ\/} {\bf 653} L53

\bibitem{amaro-seoane:2010:mbhb}
{Amaro-Seoane} P, {Eichhorn} C, {Porter} E~K and {Spurzem} R 2010 {\em MNRAS\/}
  {\bf 401} 2268

\bibitem{miller:2004:imbh}
{Miller} M~C and {Colbert} E~J~M 2004 {\em Int. J. Mod. Phys. D\/} {\bf 13} 1

\bibitem{miller:2009CQGra..26i4031M}
{Miller} M~C 2009 {\em Class.~Quantum~Grav.\/} {\bf 26} 094031

\bibitem{konstatantinidis:2011arXiv1108.5175K}
{Konstantinidis} S, {Amaro-Seoane} P and {Kokkotas} K~D 2011  ArXiv:1108.5175

\bibitem{schoedel:2003:sdca}
{Sch{\"o}del} R {\em et~al.\/} 2003 {\em ApJ\/} {\bf 596} 1015

\bibitem{ghez:2003:fmsl}
{Ghez} A~M {\em et~al.\/} 2003 {\em ApJ\/} {\bf 586} L127

\bibitem{eisenhauer:2005:sinfoni}
{Eisenhauer} F {\em et~al.\/} 2005 {\em ApJ\/} {\bf 628} 246

\bibitem{ghez:2005:sogc}
{Ghez} A~M {\em et~al.\/} 2005 {\em ApJ\/} {\bf 620} 744

\bibitem{ghez:2008:mdp}
{Ghez} A~M {\em et~al.\/} 2008 {\em ApJ\/} {\bf 689} 1044

\bibitem{gillessen:2009:mso}
{Gillessen} S {\em et~al.\/} 2009 {\em ApJ\/} {\bf 692} 1075

\bibitem{maoz:1998:dc}
{Maoz} E 1998 {\em ApJ\/} {\bf 494} L181

\bibitem{genzel:2010RvMP...82.3121G}
{Genzel} R, {Eisenhauer} F and {Gillessen} S 2010 {\em Rev. Mod. Phys.\/} {\bf
  82} 3121

\bibitem{genzel:2000MNRAS.317..348G}
{Genzel} R {\em et~al.\/} 2000 {\em MNRAS\/} {\bf 317} 348

\bibitem{genzel:2006Natur.442..786G}
{Genzel} R {\em et~al.\/} 2006 {\em Nature\/} {\bf 442} 786

\bibitem{amaro-soane:2007:tr}
{Amaro-Seoane} P {\em et~al.\/} 2007 {\em Class.~Quantum~Grav.\/} {\bf 24} 113

\bibitem{sigurdsson:1997:csm}
{Sigurdsson} S and {Rees} M~J 1997 {\em MNRAS\/} {\bf 284} 318

\bibitem{miralda-escoude:2000:bhgc}
{Miralda-Escud{\'e}} J and {Gould} A 2000 {\em ApJ\/} {\bf 545} 847

\bibitem{freitag:2006:srg}
{Freitag} M, {Amaro-Seoane} P and {Kalogera} V 2006 {\em ApJ\/} {\bf 649} 91

\bibitem{hopman:2006:rrn}
{Hopman} C and {Alexander} T 2006 {\em ApJ\/} {\bf 645} 1152

\bibitem{hopman:2006:ems}
{Hopman} C and {Alexander} T 2006 {\em ApJ\/} {\bf 645} L133

\bibitem{guerkan:2007:rr}
{G{\"u}rkan} M~A and {Hopman} C 2007 {\em MNRAS\/} {\bf 379} 1083

\bibitem{merritt:2011PhRvD..84d4024M}
{Merritt} D, {Alexander} T, {Mikkola} S and {Will} C~M 2011 {\em
  Phys.~Rev.~D\/} {\bf 84} 044024

\bibitem{miller:2005:bes}
{Miller} M~C, {Freitag} M, {Hamilton} D~P and {Lauburg} V~M 2005 {\em ApJ\/}
  {\bf 631} L117

\bibitem{brown:2009:asd}
{Brown} W~R, {Geller} M~J, {Kenyon} S~J and {Bromley} B~C 2009 {\em ApJ\/} {\bf
  690} L69

\bibitem{levin:2007:ssmbh}
{Levin} Y 2007 {\em MNRAS\/} {\bf 374} 515

\bibitem{ryan:1995:gwi}
{Ryan} F~D 1995 {\em Phys.~Rev.~D\/} {\bf 52} 5707

\bibitem{ryan:1997}
{Ryan} F~D 1997 {\em Phys. Rev. D\/} {\bf 56} 1845

\bibitem{barack:2007}
{Barack} L and {Cutler} C 2007 {\em Phys. Rev. D\/} {\bf 75} 042003

\bibitem{finn:2000:gwc}
{Finn} L~S and {Thorne} K~S 2000 {\em Phys.~Rev.~D\/} {\bf 6212} 124021

\bibitem{2006CQGra..23S.769D}
{Drasco} S 2006 {\em Classical and Quantum Gravity\/} {\bf 23} 769

\bibitem{barack:2004:lcs}
{Barack} L and {Cutler} C 2004 {\em Phys.~Rev.~D\/} {\bf 69} 082005

\bibitem{teukolsky:1973:rbh}
{Teukolsky} S~A 1973 {\em ApJ\/} {\bf 185} 635

\bibitem{gair:2009CQGra..26i4034G}
{Gair} J~R 2009 {\em Class.~Quantum~Grav.\/} {\bf 26} 094034

\bibitem{hopman:2009:emri}
{Hopman} C 2009 {\em Class.~Quantum~Grav.\/} {\bf 26} 094028

\bibitem{preto:2010:ApJ...708L..42P}
{Preto} M and {Amaro-Seoane} P 2010 {\em ApJ\/} {\bf 708} L42

\bibitem{amaro-seoane:2011CQGra..28i4017A}
{Amaro-Seoane} P and {Preto} M 2011 {\em Class.~Quantum~Grav.\/} {\bf 28}
  094017

\bibitem{Aller:2002}
{Aller} M~C and {Richstone} D 2002 {\em AJ\/} {\bf 124} 3035

\bibitem{gair:2004:ere}
{Gair} J~R {\em et~al.\/} 2004 {\em Class.~Quantum~Grav.\/} {\bf 21} S1595

\bibitem{cornish:2011CQGra..28i4016C}
{Cornish} N~J 2011 {\em Class.~Quantum~Grav.\/} {\bf 28} 094016

\bibitem{gair:2008:cmh}
{Gair} J~R, {Porter} E, {Babak} S and {Barack} L 2008 {\em
  Class.~Quantum~Grav.\/} {\bf 25} 184030

\bibitem{babak:2010:mldc}
{Babak} S {\em et~al.\/} 2010 {\em Class.~Quantum~Grav.\/} {\bf 27} 084009

\bibitem{gair:2010:emri}
{Gair} J~R, {Tang} C and {Volonteri} M 2010 {\em Phys.~Rev.~D\/} {\bf 81}
  104014

\bibitem{cornish:2006:mes}
{Cornish} N~J and {Porter} E~K 2006 {\em Class.~Quantum~Grav.\/} {\bf 23} 761

\bibitem{huerta:2009PhRvD..79h4021H}
{Huerta} E~A and {Gair} J~R 2009 {\em Phys.~Rev.~D\/} {\bf 79} 084021

\bibitem{porter:2009GWN.....1....4P}
{Porter} E~K 2009 {\em GW Notes\/} {\bf 1} 4

\bibitem{will:2006:gre}
{Will} C~M 2006 {\em Living Reviews in Relativity\/} {\bf 9} 3

\bibitem{lorimer:2008:bmp}
{Lorimer} D~R 2008 {\em Living Reviews in Relativity\/} {\bf 11} 8

\bibitem{2011PhRvD..84j1501B}
{Berti} E, {Gair} J and {Sesana} A 2011 {\em Phys. Rev. D\/} {\bf 84} 101501

\bibitem{huwyler:2011arXiv1108.1826H}
{Huwyler} C, {Klein} A and {Jetzer} P 2011  ArXiv:1108.1826

\bibitem{berti:2005:esb}
{Berti} E, {Buonanno} A and {Will} C~M 2005 {\em Phys.~Rev.~D\/} {\bf 71}
  084025

\bibitem{2012PhRvD..85f4041A}
{Alsing} J, {Berti} E, {Will} C~M and {Zaglauer} H 2012 {\em Phys. Rev. D\/}
  {\bf 85} 064041

\bibitem{yunes:2010PhRvD..81f4018Y}
{Yunes} N, {Pretorius} F and {Spergel} D 2010 {\em Phys. Rev. D\/} {\bf 81}
  064018

\bibitem{2012PhRvD..85b4041M}
{Mirshekari} S, {Yunes} N and {Will} C~M 2012 {\em Phys. Rev. D\/} {\bf 85}
  024041

\bibitem{arun:2006:pns}
{Arun} K~G, {Iyer} B~R, {Qusailah} M~S~S and {Sathyaprakash} B~S 2006 {\em
  Phys.~Rev.~D\/} {\bf 74} 024006

\bibitem{yunes:2009PhRvD..80l2003Y}
{Yunes} N and {Pretorius} F 2009 {\em Phys.~Rev.~D\/} {\bf 80} 122003

\bibitem{cornish:2011PhRvD..84f2003C}
{Cornish} N, {Sampson} L, {Yunes} N and {Pretorius} F 2011 {\em Phys.~Rev.~D\/}
  {\bf 84} 062003

\bibitem{2012PhRvD..85h2003L}
{Li} T~G~F {\em et~al.\/} 2012 {\em Phys. Rev. D\/} {\bf 85} 082003

\bibitem{rezzolla:2008ApJ...674L..29R}
{Rezzolla} L {\em et~al.\/} 2008 {\em ApJ\/} {\bf 674} L29

\bibitem{berti:2009CQGra..26p3001B}
{Berti} E, {Cardoso} V and {Starinets} A~O 2009 {\em Class.~Quantum~Grav.\/}
  {\bf 26} 163001

\bibitem{berti:2007PhRvD..76j4044B}
{Berti} E, {Cardoso} J, {Cardoso} V and {Cavagli{\`a}} M 2007 {\em
  Phys.~Rev.~D\/} {\bf 76} 104044

\bibitem{2012PhRvD..85b4018K}
{Kamaretsos} I, {Hannam} M, {Husa} S and {Sathyaprakash} B~S 2012 {\em Phys.
  Rev. D\/} {\bf 85} 024018

\bibitem{dreyer:2004:bhs}
{Dreyer} O {\em et~al.\/} 2004 {\em Class.~Quantum~Grav.\/} {\bf 21} 787

\bibitem{yoshida:1994PhRvD..50.6235Y}
{Yoshida} S, {Eriguchi} Y and {Futamase} T 1994 {\em Phys.~Rev.~D\/} {\bf 50}
  6235

\bibitem{berti:2006:boson}
Berti E and Cardoso V 2006 {\em Int. J. Mod. Phys.\/} {\bf D15} 2209

\bibitem{chirenti:2007CQGra..24.4191C}
{Chirenti} C~B~M~H and {Rezzolla} L 2007 {\em Class.~Quantum~Grav.\/} {\bf 24}
  4191

\bibitem{pani:2009PhRvD..80l4047P}
{Pani} P {\em et~al.\/} 2009 {\em Phys.~Rev.~D\/} {\bf 80} 124047

\bibitem{barausse:2008xv}
Barausse E and Sotiriou T~P 2008 {\em Phys. Rev. Lett.\/} {\bf 101} 099001

\bibitem{pani:2009PhRvD..79h4031P}
{Pani} P and {Cardoso} V 2009 {\em Phys.~Rev.~D\/} {\bf 79} 084031

\bibitem{barausse:2007PhRvD..75f4026B}
{Barausse} E, {Rezzolla} L, {Petroff} D and {Ansorg} M 2007 {\em
  Phys.~Rev.~D\/} {\bf 75} 064026

\bibitem{barausse:2008PhRvD..77j4027B}
{Barausse} E and {Rezzolla} L 2008 {\em Phys.~Rev.~D\/} {\bf 77} 104027

\bibitem{kocsis:2011PhRvD..84b4032K}
{Kocsis} B, {Yunes} N and {Loeb} A 2011 {\em Phys.~Rev.~D\/} {\bf 84} 024032

\bibitem{yunes:2011PhRvD..83d4030Y}
{Yunes} N, {Coleman Miller} M and {Thornburg} J 2011 {\em Phys.~Rev.~D\/} {\bf
  83} 044030

\bibitem{2012ApJ...744L..20A}
{Amaro-Seoane} P, {Brem} P, {Cuadra} J and {Armitage} P~J 2012 {\em ApJL\/}
  {\bf 744} L20

\bibitem{drasco:2004:rbh}
{Drasco} S and {Hughes} S~A 2004 {\em Phys.~Rev.~D\/} {\bf 69} 044015

\bibitem{sopuerta:2010:emri}
{Sopuerta} C~F 2010 {\em GW Notes\/} {\bf 4} 3

\bibitem{babak:2011CQGra..28k4001B}
{Babak} S, {Gair} J~R, {Petiteau} A and {Sesana} A 2011 {\em
  Class.~Quantum~Grav.\/} {\bf 28} 114001

\bibitem{kesden:2005:gws}
{Kesden} M, {Gair} J and {Kamionkowski} M 2005 {\em Phys.~Rev.~D\/} {\bf 71}
  044015

\bibitem{gair:2008:pbh}
{Gair} J~R 2009 {\em Class.~Quantum~Grav.\/} {\bf 26} 094034

\bibitem{lukes-gerakopoulos:2010PhRvD..81l4005L}
{Lukes-Gerakopoulos} G, {Apostolatos} T~A and {Contopoulos} G 2010 {\em
  Phys.~Rev.~D\/} {\bf 81} 124005

\bibitem{collins:2004:tfm}
{Collins} N~A and {Hughes} S~A 2004 {\em Phys.~Rev.~D\/} {\bf 69} 124022

\bibitem{glampedakis:2006:msl}
{Glampedakis} K and {Babak} S 2006 {\em Class.~Quantum~Grav.\/} {\bf 23} 4167

\bibitem{vigeland:2011PhRvD..83j4027V}
{Vigeland} S, {Yunes} N and {Stein} L~C 2011 {\em Phys.~Rev.~D\/} {\bf 83}
  104027

\bibitem{sopuerta:2009PhRvD..80f4006S}
{Sopuerta} C~F and {Yunes} N 2009 {\em Phys.~Rev.~D\/} {\bf 80} 064006

\bibitem{pani:2011PhRvD..83j4048P}
{Pani} P, {Cardoso} V and {Gualtieri} L 2011 {\em Phys.~Rev.~D\/} {\bf 83}
  104048

\bibitem{yagi:2010PhRvD..81f4008Y}
{Yagi} K and {Tanaka} T 2010 {\em Phys.~Rev.~D\/} {\bf 81} 064008

\bibitem{mcwilliams:2010PhRvL.104n1601M}
{McWilliams} S~T 2010 {\em Phys.~Rev.~Lett.\/} {\bf 104} 141601

\bibitem{yagi:2011PhRvD..83h4036Y}
{Yagi} K, {Tanahashi} N and {Tanaka} T 2011 {\em Phys.~Rev.~D\/} {\bf 83}
  084036

\bibitem{Cardoso:2011xi}
Cardoso V {\em et~al.\/} 2011 {\em Phys. Rev. Lett.\/} {\bf 107} 241101

\bibitem{2012PhRvD..85j2003Y}
{Yunes} N, {Pani} P and {Cardoso} V 2012 {\em Phys. Rev. D\/} {\bf 85} 102003

\bibitem{gair:2011PhRvD..84f4016G}
{Gair} J and {Yunes} N 2011 {\em Phys.~Rev.~D\/} {\bf 84} 064016

\bibitem{2012JCAP...06..027B}
{Bin{\'e}truy} P, {Boh{\'e}} A, {Caprini} C and {Dufaux} J~F 2012 {\em J.
  Cosmology Astropart. Phys.\/} {\bf 6} 027

\bibitem{witten:1984:csp}
{Witten} E 1984 {\em Phys.~Rev.~D\/} {\bf 30} 272

\bibitem{hogan:1986:grc}
{Hogan} C~J 1986 {\em MNRAS\/} {\bf 218} 629

\bibitem{Kamionkowski:1993fg}
Kamionkowski M, Kosowsky A and Turner M~S 1994 {\em Phys. Rev.\/} {\bf D49}
  2837

\bibitem{Huber:2008hg}
Huber S~J and Konstandin T 2008 {\em JCAP\/} {\bf 0809} 022

\bibitem{Caprini:2009yp}
Caprini C, Durrer R and Servant G 2009 {\em JCAP\/} {\bf 0912} 024

\bibitem{hogan:2000:gwm}
{Hogan} C~J 2000 {\em Phys.~Rev.~Lett.\/} {\bf 85} 2044

\bibitem{randall:2006:gww}
{Randall} L and {Servant} G 2007 {\em JHEP\/} {\bf 5} 54

\bibitem{copeland:2004JHEP...06..013C}
{Copeland} E~J, {Myers} R~C and {Polchinski} J 2004 {\em JHEP\/} {\bf 6} 13

\bibitem{1994csot.book.....V}
{Vilenkin} A and {Shellard} E~P~S 1994 {\em {Cosmic strings and other
  topological defects}\/} (Cambridge, UK: Princeton University Press)

\bibitem{khlebnikov:1997:rgw}
{Khlebnikov} S and {Tkachev} I 1997 {\em Phys.~Rev.~D\/} {\bf 56} 653

\bibitem{easther:2006:sgw}
{Easther} R and {Lim} E~A 2006 {\em J. Cosmology Astropart. Phys.\/} {\bf 4} 10

\bibitem{GarciaBellido:2007dg}
Garcia-Bellido J and Figueroa D~G 2007 {\em Phys. Rev. Lett.\/} {\bf 98} 061302

\bibitem{Dufaux:2007pt}
Dufaux J~F {\em et~al.\/} 2007 {\em Phys. Rev.\/} {\bf D76} 123517

\bibitem{Dufaux:2008dn}
Dufaux J~F, Felder G, Kofman L and Navros O 2009 {\em JCAP\/} {\bf 0903} 001

\bibitem{brustein:1995:rgw}
{Brustein} R, {Gasperini} M, {Giovannini} M and {Veneziano} G 1995 {\em Phys.
  Lett. B\/} {\bf 361} 45

\bibitem{buonanno:2003:tasi}
{Buonanno} A 2003  {gr-qc/0303085}

\bibitem{buonanno:1997:srg}
{Buonanno} A, {Maggiore} M and {Ungarelli} C 1997 {\em Phys.~Rev.~D\/} {\bf 55}
  3330

\bibitem{hogan:2001:esg}
{Hogan} C~J and {Bender} P~L 2001 {\em Phys.~Rev.~D\/} {\bf 64} 062002

\bibitem{damour:2005:grc}
{Damour} T and {Vilenkin} A 2005 {\em Phys.~Rev.~D\/} {\bf 71} 063510

\bibitem{siemens:2006:gwb}
{Siemens} X {\em et~al.\/} 2006 {\em Phys.~Rev.~D\/} {\bf 73} 105001

\bibitem{binetruy:2010PhRvD..82l6007B}
{Bin{\'e}truy} P, {Boh{\'e}} A, {Hertog} T and {Steer} D~A 2010 {\em
  Phys.~Rev.~D\/} {\bf 82} 126007

\bibitem{bohe:2011PhRvD..84f5016B}
{Boh{\'e}} A 2011 {\em Phys.~Rev.~D\/} {\bf 84} 065016

\bibitem{schutz:1986:hgw}
{Schutz} B~F 1986 {\em Nature\/} {\bf 323} 310

\bibitem{holz:2005:ugw}
{Holz} D~E and {Hughes} S~A 2005 {\em ApJ\/} {\bf 629} 15

\bibitem{armitage:2002ApJ...567L...9A}
{Armitage} P~J and {Natarajan} P 2002 {\em ApJ\/} {\bf 567} L9

\bibitem{milosavljevic:2005:amb}
{Milosavljevi{\'c}} M and {Phinney} E~S 2005 {\em ApJ\/} {\bf 622} L93

\bibitem{phinney:2009astro2010S.235P}
{Phinney} E~S 2009  ArXiv:0903.0098

\bibitem{schnittman:2011CQGra..28i4021S}
{Schnittman} J~D 2011 {\em Class.~Quantum~Grav.\/} {\bf 28} 094021

\bibitem{lsst:2009arXiv0912.0201L}
{LSST Science Collaborations} 2009  ArXiv:0912.0201

\bibitem{johnston:2007PASA...24..174J}
{Johnston} S {\em et~al.\/} 2007 {\em PASA\/} {\bf 24} 174

\bibitem{riess:1998ApJ...504..935R}
{Riess} A~G~o 1998 {\em ApJ\/} {\bf 504} 935

\bibitem{perlmutter:1999AIPC..478..129P}
{Perlmutter} S and {Riess} A 1999 {\em COSMO-98\/} ({\em AIP~Conf.~Series\/}
  vol 478) ed {D~O~Caldwell} p 129

\bibitem{petiteau:2011ApJ...732...82P}
{Petiteau} A, {Babak} S and {Sesana} A 2011 {\em ApJ\/} {\bf 732} 82

\bibitem{macleod:2008:phc}
{MacLeod} C~L and {Hogan} C~J 2008 {\em Phys.~Rev.~D\/} {\bf 77} 043512

\end{thebibliography}

\end{document}